\documentclass[twocolumn,
prd,aps,superscriptaddress,tightenlines,floats,nofootinbib]{revtex4}
\usepackage{graphicx}
\usepackage{bm}
\usepackage{axodraw}
\def\nn{\nonumber\\}

\begin{document}

\title{Baryon Magnetic Moments in the $1/N_c$ Expansion with Flavor Symmetry Breaking}

\author{Elizabeth Jenkins}
\affiliation{Department of Physics, University of California at San Diego,
  La Jolla, CA 92093\vspace{4pt} }
\date{\today}

\vskip1.5in
\begin{abstract}
The magnetic moments and transition magnetic moments of the ground state baryons are analyzed in an
expansion in $1/N_c$, $SU(3)$ flavor symmetry breaking and isospin symmetry breaking.  There is clear evidence in the experimental data for the hierarchy of magnetic moments of the combined expansion in $1/N_c$ and flavor breaking.  $SU(3)$ breaking in the magnetic moments is expected to be enhanced relative to that of other hadronic observables, and significant $SU(3)$ breaking is found.
\end{abstract}

\maketitle
\tableofcontents

\section{Introduction}

The magnetic moments of the ground state baryons have been of experimental and theoretical interest for over 50 years.  The baryon magnetic moments have been studied using a variety of methods, some of which predate QCD and helped lead to the formulation of the theory, including $SU(3)$ flavor symmetry, $SU(3)$ chiral perturbation theory and the $1/N_c$ expansion.  This work will consider the baryon magnetic moments in a combined expansion in $1/N_c$ and flavor symmetry breaking, both $SU(3)$ flavor symmetry breaking and isospin breaking. 
To date, a complete analysis in a combined expansion in $1/N_c$ and $SU(3)$ flavor symmetry breaking has not been done.  
Analyses in flavor symmetry breaking (i.e.\ chiral perturbation theory) alone have been performed in 
works~\cite{Caldi,Gasser,Krause,JLMS,Butler}.  Analyses of the baryon magnetic moments in the $1/N_c$ expansion occur in 
Refs.~\cite{djmone,lmw,jm,djmtwo,ddjm,lebedmartin,fmone,fmtwo}.  
The main result of this work is that the combined expansion in $1/N_c$ and $SU(3)$ breaking is needed to understand the hierarchy of baryon magnetic moments found in nature.  The derivation given here of the combined expansion to all orders in $SU(3)$ breaking and in $1/N_c$ is analogous to the analysis of the baryon masses of Ref.~\cite{jl}.     

The theoretical analysis performed in this work is motivated by new experimental results.  Recently, first measurements of two additional decuplet-octet baryon transition magnetic moments have appeared~\cite{CLASone,CLAStwo}.
The new experimental data as well as the promise of future measurements makes it interesting to address the question of the pattern of baryon magnetic moments more completely than in previous work.

The organization of this paper is as follows.  Section~II analyzes the baryon magnetic moments in the $1/N_c$ expansion in the $SU(3)$ flavor symmetry limit.  Four $1/N_c$ operators transforming as spin-$1$ flavor-octets under the $SU(2) \otimes SU(3)$ spin $\times$ flavor group parametrize the 27 baryon magnetic moments in the $SU(3)$ flavor symmetry limit.  Thus, 23 baryon magnetic moment relations are implied by exact $SU(3)$ symmetry.  Section~III contains the $SU(3)$ flavor symmetry breaking analysis of the baryon magnetic moments to all orders in the $1/N_c$ expansion.  The complete set of 27 $1/N_c$ operators transforming according to definite $SU(3)$ representations is determined.  In this operator basis, the $SU(3)$ structure of the baryon magnetic moments is manifest.   
Section~IV contains the main results of this work.  A hierarchy of magnetic moment relations is obtained in the combined $1/N_c$ and flavor-symmetry breaking expansion.  Since many of the ground state baryon magnetic moments are not measured, most of the linear combinations of magnetic moments cannot be evaluated from the experimental data at this time.  However, it is possible to perform a variety of fits to the experimental data at successive orders in the combined expansion.
Clear evidence in the experimental data for the combined $1/N_c$ and $SU(3)$ symmetry breaking expansion is obtained from these fits.
The overall spin-flavor symmetry breaking structure of the baryon magnetic moments is found to be an intricate pattern of $1/N_c$ and $SU(3)$ flavor-symmetry breaking suppressions.  It is possible to evaluate all 27 baryon magnetic moments in terms of the leading coefficients extracted from each fit to the experimental data, yielding predictions for the many unmeasured baryon magnetic moments.  
The complete $SU(3)$ flavor symmetry breaking pattern of the baryon magnetic moments
also is explicit in the analysis presented here.  Known $SU(3)$-violating relations for the baryon octet magnetic moments~\cite{Caldi,JLMS} are rederived, and corresponding relations for the decuplet magnetic moments and the decuplet-octet transition magnetic moments also are obtained.  The $SU(3)$ flavor symmetry breaking of the baryon magnetic moments is predicted on the basis of $SU(3)$ chiral perturbation theory to be dominated by chiral loop corrections which are nonanalytic in the light quark masses.  Evidence for the group theory of this flavor symmetry breaking pattern is found.  
Conclusions and final remarks are given in Section~V.

\section{$SU(3)$ Flavor Symmetry Analysis}

The 27 magnetic moments of the ground state baryons consist of nine magnetic moments of the baryon octet, the eight magnetic moments of the individual octet baryons $p$, $n$, etc. and the $\Sigma^0 \rightarrow \Lambda$ transition magnetic moment $\Lambda \Sigma^0$, ten magnetic moments of the individual decuplet baryons, and eight transition magnetic moments of a decuplet baryon to an octet baryon.  
For the flavor-symmetry breaking analysis, it is useful to group the magnetic moments in terms of linear combinations with $I=0,1,2$ and $3$.  There are ten $I=0$ combinations,  
\begin{eqnarray}\label{isozero}
N_0 &\equiv& {1 \over 2} \left( p + n \right), \nn
\Lambda_0 &\equiv& \Lambda,\nn
\Sigma_0 &\equiv& {1 \over 3} \left( \Sigma^+ + \Sigma^0 + \Sigma^- \right), \nn
\Xi_0 &\equiv& {1 \over 2} \left( \Xi^0 + \Xi^- \right), \nn
\Delta_0 &\equiv& {1 \over 4} \left( \Delta^{++} + \Delta^+ + \Delta^0 + \Delta^- \right), \nn
\Sigma_0^* &=& {1\over 3} \left( \Sigma^{*+} + \Sigma^{*0} + \Sigma^{*-} \right), \nn
\Xi^*_0 &=& {1 \over 2} \left( \Xi^{*0} + \Xi^{*-} \right), \nn
\Omega_0 &\equiv& \Omega^-, \nn
\left( \Sigma \Sigma^*\right)_0 &\equiv& {1 \over 3}\left( \Sigma\Sigma^{*+} + \Sigma \Sigma^{*0} + \Sigma \Sigma^{*-} \right), \nn
\left( \Xi \Xi^* \right)_0 &\equiv& {1 \over 2} \left( \Xi \Xi^{*0} + \Xi \Xi^{*-} \right),
\end{eqnarray}
eleven $I=1$ combinations,
\begin{eqnarray}\label{isoone}
N_1 &\equiv& \left( p-n \right), \nn
\left( \Lambda \Sigma \right)_1 &\equiv& \Lambda \Sigma^0, \nn
\Sigma_1 &\equiv& \left( \Sigma^+ - \Sigma^- \right), \nn
\Xi_1 &\equiv& \left( \Xi^0 - \Xi^- \right), \nn
\Delta_1 &\equiv& \left( 3 \Delta^{++} + \Delta^+ -\Delta^0 -3 \Delta^- \right), \nn
\Sigma^*_1 & \equiv& \left( \Sigma^{*+} - \Sigma^{*-} \right), \nn
\Xi_1^* &\equiv& \left( \Xi^{*0} - \Xi^{*-} \right), \nn
\left( N \Delta \right)_1 &\equiv& \left( p\Delta^+ + n\Delta^0 \right), \nn
\left( \Lambda \Sigma^* \right)_1 &\equiv& \Lambda \Sigma^{*0}, \nn
\left( \Sigma \Sigma^* \right)_1 &\equiv& \left( \Sigma \Sigma^{*+} - \Sigma \Sigma^{*-} \right), \nn
\left( \Xi \Xi^* \right)_1 &\equiv& \left( \Xi \Xi^{*0} - \Xi \Xi^{*-} \right),
\end{eqnarray}
five $I=2$ combinations
\begin{eqnarray}\label{isotwo}
\Sigma_2 &\equiv& \left( \Sigma^+ -2 \Sigma^0 + \Sigma^- \right), \nn
\Delta_2 &\equiv& \left( \Delta^{++} - \Delta^{+} - \Delta^0 + \Delta^- \right), \nn
\Sigma^*_2 &\equiv& \left( \Sigma^{*+} -2 \Sigma^{*0} + \Sigma^{*-} \right), \nn
\left( N \Delta \right)_2 &\equiv& \left( p \Delta^+ - n \Delta^0 \right), \nn
\left( \Sigma \Sigma^* \right)_2 &\equiv& \left( \Sigma \Sigma^{*+} -2 \Sigma \Sigma^{*0} + \Sigma \Sigma^{*-}\right),
\end{eqnarray}
and one $I=3$ combination, 
\begin{eqnarray}\label{isothree}
\Delta_3 &\equiv& \left( \Delta^{++} -3 \Delta^+ + 3 \Delta^0 -\Delta^-\right) ,
\end{eqnarray}
where the notation used for the transition magnetic moments is the same as Ref.~\cite{jm}.
In the limit of exact isospin symmetry, all magnetic moment operators are linear combinations of isovector $I=1$ and isoscalar $I=0$ operators,
and so the five $I=2$ magnetic moment combinations $\Sigma_2$, $\Delta_2$, $\Sigma^*_2$, $\left( N \Delta \right)_2$ and 
$\left( \Sigma \Sigma^* \right)_2$ and the one $I=3$ magnetic moment combination $\Delta_3$ vanish.  
The general analysis of this work includes isospin symmetry breaking, allowing these six isospin-violating combinations to be nonzero.  However, in much of the analysis, isospin symmetry breaking will be neglected.  When isospin breaking is neglected, the
six $I=2$ and $I=3$ isospin-violating magnetic moment combinations give relations which can be used to determine six magnetic moments in terms of the others.  As a consequence, there are only 21 independent magnetic moments in the isospin symmetry
limit.  Without loss of generality, the six dependent magnetic moments in the isospin symmetry limit are taken to be
\begin{eqnarray}\label{isospin}
n \Delta^0 &=& p \Delta^+, \nn
\Delta^+ &=& {1 \over 3} \left( 2 \Delta^{++} + \Delta^- \right), \nn
\Delta^0 &=& {1 \over 3} \left( \Delta^{++} + 2 \Delta^- \right), \nn
\Sigma^0 &=& {1 \over 2} \left( \Sigma^+ + \Sigma^- \right), \nn
\Sigma^{*0} &=& {1 \over 2} \left( \Sigma^{*+} + \Sigma^{*-} \right), \nn
\Sigma\Sigma^{*0} &=& {1 \over 2} \left( \Sigma\Sigma^{*+} + \Sigma\Sigma^{*-} \right).
\end{eqnarray}

Progress in understanding the spin-flavor structure of static baryon properties has resulted from the discovery of a contracted spin-flavor symmetry for baryons in the large $N_c$ limit, where $N_c$ is the number of colors in 
QCD~\cite{dmone,jone,dmtwo,jtwo,jthree,djmone}.  Expansion in $1/N_c$ about the large-$N_c$ limit~\cite{thooft,witten} has
led to a classification of baryon operators in $1/N_c$ and flavor symmetry 
breaking~\cite{dmone,jone,dmtwo,jtwo,jthree,djmone,cgo,lmr,jm,djmtwo,lmw,jl,ddjm,j,lebedmartin,JMNW,fmone,fmtwo}.  A wide variety of baryon properties have been studied.
It has been shown that the $1/N_c$ expansion gives a good description of the spin-flavor structure of QCD baryons with $N_c=3$.

The form of the baryon magnetic moment $1/N_c$ expansion is
\begin{eqnarray}\label{expn}
\sum_{n=1}^{N_c } c_{(n)} \ {1 \over N_c^{n-1}} {\cal O}_{(n)} ,
\end{eqnarray}
where each independent $n$-body operator ${\cal O}_{(n)}$ in the expansion is 
a product of $n$ of the baryon spin-flavor generators
$J^i$, $T^a$ and $G^{jb}$~\cite{djmtwo,lmr}.    
$1/N_c$ power counting rules dictate the explicit factor of $1/N_c^{n-1}$ accompanying each 
$n$-body operator ${\cal O}_{(n)}$ in the expansion.  
The $N_c$-dependence of each term in the $1/N_c$ expansion is
a product of this explicit factor times the implicit $N_c$-dependence of matrix elements of the operator ${\cal O}_{(n)}$.
Matrix elements of the spin generators $J^i$ are $O(1)$, so operators with additional powers of the spin operator generally 
are suppressed in the $1/N_c$ expansion. Matrix elements of $T^a$ and 
$G^{ia}$ have $N_c$-dependence which varies in different portions of the flavor weight diagrams of large-$N_c$ baryons and for different values of $a=1, \cdots,8$, making the analysis more subtle.  Each independent operator in the $1/N_c$ expansion is accompanied by an unknown coefficient $c_{(n)}$ which is order unity at leading order in the $1/N_c$ expansion.  Predictions based on the $1/N_c$ expansion are obtained by neglecting operators which are suppressed by powers of $1/N_c$. 
For baryons with finite $N_c$,
the baryon $1/N_c$ expansion extends only up to $N_c$-body operators.  Thus,  
for QCD baryons with $N_c=3$, the $1/N_c$ expansion of Eq.~(\ref{expn})
ends with $3$-body operators.

In the $SU(3)$ flavor symmetry limit, the baryon magnetic moment operator transforms as a $(1, {\bf 8})$ representation under $SU(2) \otimes SU(3)$ spin $\times$ flavor symmetry.  For QCD baryons with $N_c=3$, the $1/N_c$ expansion for the baryon magnetic moments is~\cite{djmtwo} 
\begin{eqnarray}\label{su3}
M^{iQ} &=& a^{1,{\bf 8}}_{(1)}\ G^{iQ} + b^{1,{\bf 8}}_{(2)}\ {1 \over N_c} J^i T^Q  \nn
&&+ b^{1,{\bf 8}}_{(3)}\ {1 \over N_c^2} {\cal D}_{(3)}^{iQ} 
+ c^{1,{\bf 8}}_{(3)}\ {1 \over N_c^2}{\cal O}_{(3)}^{iQ},
\end{eqnarray}
where 
\begin{eqnarray}
T^Q &=& \pmatrix{ {2 \over 3} & 0 & 0 \cr
0 & -{1 \over 3} & 0 \cr
0 & 0 & -{1 \over 3} \cr} = T^3 + {1 \over \sqrt{3}} T^8 = Q
\end{eqnarray}
is the $SU(3)$ flavor generator corresponding to the light quark electromagnetic charge,
and
\begin{eqnarray}\label{geq}
G^{iQ}  &\equiv& G^{i3} + {1 \over \sqrt{3}} G^{i8},
\end{eqnarray}
is the spin-flavor generator in the $Q$ flavor direction.
The two $3$-body operators in the $1/N_c$ expansion are defined by
\begin{eqnarray}
{\cal D}_{(3)}^{iQ} &=& \{ J^i, \{ J^j, G^{jQ} \} \}, \nn
{\cal O}_{(3)}^{iQ} &=& \{ J^2, G^{iQ} \} - {1 \over 2}\{ J^i, \{ J^j, G^{jQ} \} \}.
\end{eqnarray}
The $2$-body operator $J^i T^Q$ and the $3$-body operator
${\cal D}_{(3)}^{iQ}$ are purely diagonal operators with no nonvanishing matrix elements between decuplet and octet baryons.  The $3$-body operator  
${\cal O}_{(3)}^{iQ}$ is purely off-diagonal; its only nonvanishing matrix elements are decuplet-octet transition matrix elements.  In contrast, the $1$-body operator $G^{iQ}$ has both diagonal and off-diagonal matrix elements.  
This $1$-body operator is the leading operator of the $1/N_c$ expansion, since the matrix elements of $G^{i3}$ are $O(N_c)$
for large-$N_c$ baryons.  The $G^{iQ}$ operator also receives a subleading $O(1)$ contribution from the matrix elements of
\begin{eqnarray}
G^{i8} &\equiv& {1 \over \sqrt{3}} \left( J^i - 3 J_s^i \right).
\end{eqnarray}
The three subleading operators $J^i T^Q$, ${\cal D}_{(3)}^{iQ}$ and ${\cal O}_{(3)}^{iQ}$ each contribute at $O(1/N_c)$.
This statement requires explanation.  The matrix elements of $J^i T^3$ are $O(1)$, so $J^i T^3/N_c$ matrix elements are 
$O(1/N_c)$.  
The operator
\begin{eqnarray}
T^8 &\equiv& {1 \over \sqrt{3}} \left( N_c - 3 N_s \right),
\end{eqnarray}
nominally has a piece of $O(N_c)$, but its contribution to magnetic moment splittings only comes from the term proportional to 
$N_s$, so 
$J^i T^8/N_c$ also contributes to magnetic moment splittings at $O(1/N_c)$.  The matrix elements of the $3$-body operators 
${\cal D}_{(3)}^{iQ}/N_c^2$ and ${\cal O}_{(3)}^{iQ}/N_c^2$ are suppressed by $1/N_c^2$ relative to the leading operator 
$G^{iQ}$, and so are $O(1/N_c)$.

The matrix elements of the four $1/N_c$ operators in Eq.~(\ref{su3}) are tabulated in the first four columns of Table~I.  The matrix elements given in the table are for each operator with $i=3$.  For octet baryons, the matrix elements are taken between octet baryons with $J_z= 1/2$, whereas for decuplet baryons, the matrix elements are given between decuplet baryons with $J_z = 3/2$.  For transition magnetic moments between decuplet and octet baryons, the matrix elements are given between $J_z = 1/2$ states for both the decuplet and octet baryons.  This same convention for the matrix elements was used in Ref.~\cite{jm}.   
Eq.~(\ref{su3}) implies that each of the 27 baryon magnetic moments is parametrized by four $1/N_c$ operator coefficients in the 
$SU(3)$ flavor symmetry limit.  For example, 
\begin{eqnarray}\label{example}
p &=& {1 \over 2} a_{(1)}^{1, {\bf 8}} + {1 \over 2} {1 \over N_c} b_{(2)}^{1, {\bf 8}} + {3 \over 2} {1 \over N_c^2} b_{(3)}^{1, {\bf 8}}, \nn
n &=& -{1 \over 3}a_{(1)}^{1, {\bf 8}} - { 1\over N_c^2} b_{(3)}^{1, {\bf 8}} , \nn
\Delta^+ &=& {1 \over 2} a_{(1)}^{1, {\bf 8}}  + {3 \over 2} {1 \over N_c} b_{(2)}^{1, {\bf 8}}
+ {{15} \over 2} {1 \over N_c^2} b_{(3)}^{1, {\bf 8}}, \nn
{1 \over \sqrt{2}} p\Delta^+ &=& {1 \over 3} a_{(1)}^{1, {\bf 8}} + {3 \over 2} {1 \over N_c^2} c_{(3)}^{1, {\bf 8}},
\end{eqnarray}
for the strangeness zero baryons.  For QCD baryons, $N_c=3$, so it is to be understood that $N_c=3$ in the above equation.

In the $SU(3)$ symmetry limit, the 27 baryon magnetic moments are given in terms of 4 parameters, so there are 23 $SU(3)$ relations for the baryon magnetic moments.  
The nine magnetic moments of the octet baryons are described in terms of only two linear combinations of the four $1/N_c$ coefficients.  Thus, there are seven $SU(3)$ relations for the octet baryon magnetic moments, the Coleman-Glashow relations~\cite{cg},
\begin{eqnarray}\label{su3octet}
p &=& \Sigma^+,  \nn
\Sigma^- &=& \Xi^-,   \nn
n &=& 2 \Lambda = \Xi^0 = {2 \over \sqrt{3}} \Lambda \Sigma^0 \nn
&=& -\left( \Sigma^+ + \Sigma^-\right) =  -2\Sigma^0 .
\end{eqnarray}
The ten magnetic moments of the decuplet baryons are given in terms of one linear combination of the four $1/N_c$ coefficients, so there are nine $SU(3)$ relations for the decuplet baryon magnetic moments,
\begin{eqnarray}\label{su3decup}
{1 \over 2} \Delta^{++} &=& \Delta^+ = -\Delta^- = \Sigma^{*+}=  -\Sigma^{*-}= -\Xi^{*-} = -\Omega^- , \nn
\Delta^0 &=& \Sigma^{*0} = \Xi^{*0} = 0.
\end{eqnarray}
The eight transition magnetic moments between decuplet and octet baryons also are given in terms of one linear combination of the four $1/N_c$ operator coefficients, yielding seven $SU(3)$ relations for the decuplet-octet transition magnetic moments
\begin{eqnarray}\label{su3decupletoctet}
p\Delta^+ &=& n\Delta^0 = \Sigma\Sigma^{*+}=\Xi\Xi^{*0} = {2 \over \sqrt{3}} \Lambda \Sigma^{*0} = 2 \Sigma\Sigma^{*0}, \nn
\Sigma\Sigma^{*-} &=& \Xi\Xi^{*-} = 0.
\end{eqnarray}
Eqs.~(\ref{su3octet}), (\ref{su3decup}) and (\ref{su3decupletoctet}) are the 23 $SU(3)$ relations for the magnetic moments.
Note that in the $SU(3)$ limit, five baryon magnetic moments vanish.
In addition, six of the above 23 $SU(3)$ relations correspond to the six isospin relations Eq.~(\ref{isospin}).

\begin{figure}
\includegraphics[]{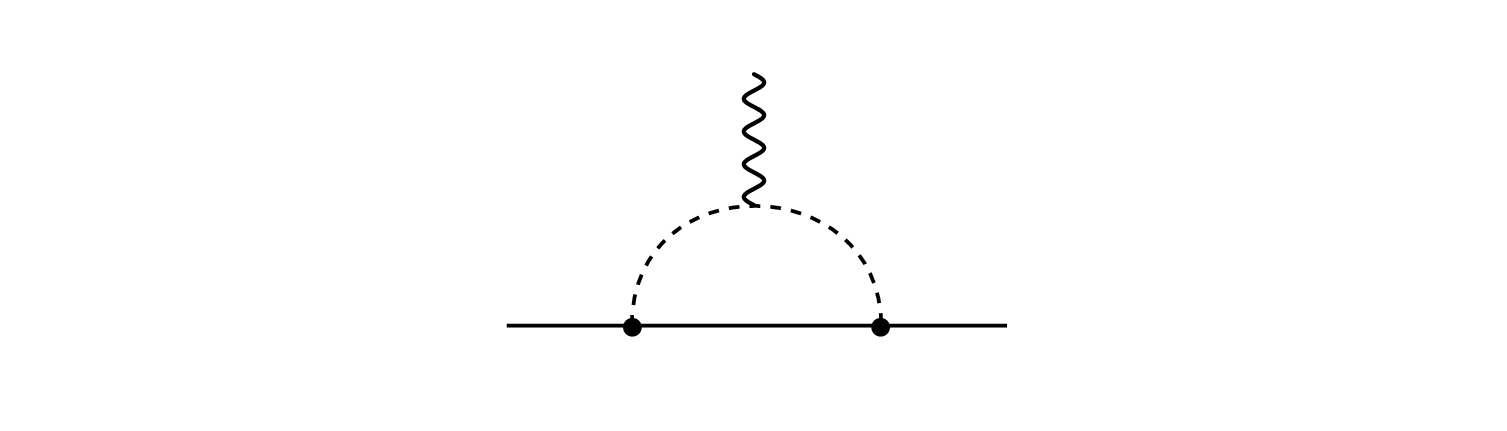}
\caption{\label{fig:1} One loop correction to the baryon magnetic moments of order $m_q^{1/2}$ for $\Delta M=0$.}
\end{figure}

The level of accuracy at which the $SU(3)$ relations are satisfied gives a quantitative measure of the size of $SU(3)$ breaking for the baryon magnetic moments.  Writing each $SU(3)$ relation as ${\rm LHS}= {\rm RHS}$ such that all terms on the left- and right-hand sides of the equation contribute with the same sign, the accuracy of an $SU(3)$ relation is defined to be $|{\rm LHS}- {\rm RHS} |/ |{\rm LHS} + {\rm RHS}|/2 $.  There are nine $SU(3)$ relations which can be tested by present experimental data.  The experimental accuracies of these nine relations are
\begin{eqnarray}\label{su3accuracies}
\begin{array}{lr}
p = \Sigma^+, &  12.8 \pm 0.4\% \\
\Sigma^- = \Xi^-, & 56.3 \pm 2.9 \% \\
n = 2 \Lambda, & 43.8 \pm 0.5 \% \\
n = \Xi^0, & 41.9 \pm 0.5 \% \\
n =  {2 \over \sqrt{3}} \Lambda \Sigma^0, & 2.8 \pm 4.8 \% \\
n = -\left( \Sigma^+ + \Sigma^-\right), & 38.3 \pm 1.7 \% \\
{1 \over 2} \Delta^{++} = -\Omega^- , & 41.4 \pm 10.5 \% \\
p\Delta^+ = \Lambda \Sigma^{*0}, & 24.3 \pm 8.6 \% \\
p\Delta^+ = \Sigma \Sigma^{*+},  & 8.6 \pm 10.7 \% \\
\end{array}
\end{eqnarray}
where the experimental data is taken from the Particle Data Group~\cite{pdg} and the two CLAS measurements~\cite{CLASone,CLAStwo}.
The experimental accuracies of Eq.~(\ref{su3accuracies}) reveal that $SU(3)$ breaking is quite sizeable for the baryon magnetic moments; it is larger than the canonical $30 \%$ breaking found for many other hadronic observables.  $SU(3)$ chiral perturbation theory predicts a definite pattern of $SU(3)$ breaking, with the leading $SU(3)$ flavor breaking arising from a one-loop chiral correction which is proportional to $m_q^{1/2}$ from the graph in Fig.~\ref{fig:1}.   Thus, one naively expects enhanced $SU(3)$ breaking of order $\sqrt{1/3} \sim 60 \%$ for the baryon magnetic moments.   $SU(3)$ breaking as large as this is seen in the experimental accuracies of Eq.~(\ref{su3accuracies}).

One can obtain a quantitative estimate of the four $1/N_c$ coefficients appearing in the $1/N_c$ expansion Eq.~(\ref{su3}) for the baryon magnetic moments in the $SU(3)$ symmetry limit.
Taking the
independent magnetic moments to be the strangeness zero baryon magnetic moments $p$, $n$, $\Delta^{++}$, $p\Delta^+$, one finds that the four $1/N_c$ coefficients are given by the linear combinations
\begin{eqnarray}\label{su3coeffs}
a_{(1)}^{1, {\bf 8}} &=&  -p -4n +{1 \over 6} \Delta^{++} ,\nn
{1 \over N_c} b_{(2)}^{1, {\bf 8}} &=& 2 p +3n ,\nn
{1 \over N_c^2} b_{(3)}^{1, {\bf 8}} &=&  -{1 \over 2}(p+ n) +{1 \over {12}}\Delta^{++} ,\nn
{1 \over N_c^2} c_{(3)}^{1, {\bf 8}} &=& -{1 \over {3}}(p-n) + {1 \over {18}} \Delta^{++}+ {2 \over 3} {1 \over \sqrt{2}} p \Delta^+  ,
\end{eqnarray}
which yields
\begin{eqnarray}\label{abbccoeffs}
a_{(1)}^{1, {\bf 8}} &=&  5.88 \pm  0.09,\nn
b_{(2)}^{1, {\bf 8}} &=& -0.459 \pm 0.000, \nn
b_{(3)}^{1, {\bf 8}} &=& 0.63 \pm 0.36, \nn
c_{(3)}^{1, {\bf 8}} &=& 3.87 \pm 0.45 \ .
\end{eqnarray}
These values for the four coefficients parametrizing the $SU(3)$ symmetry limit will be useful for comparison with the values extracted from fits to experimental data given in Section~IV.

The analysis of the $N_c$-dependence of the operator matrix elements implies that the $1/N_c$ expansion of the magnetic moment operator in the $SU(3)$ limit can be truncated after the first operator $G^{iQ}$ up to corrections of order $1/N_c$.
The $1/N_c$ expansion, 
\begin{eqnarray}\label{1ops}
M^{iQ} &=& a^{1,{\bf 8}}_{(1)}\ G^{iQ} + O\left({1 \over N_c}\right),
\end{eqnarray}
parametrizes the 27 magnetic moments in terms of one $1/N_c$ coefficient, so there are three additional $SU(3)$ relations when the subleading $O(1/N_c)$ terms in the $1/N_c$ expansion are neglected.
The three additional relations, holding in the $SU(3)$ limit at $O(1)$ in the $1/N_c$ expansion, are
\begin{eqnarray}\label{su6}
\begin{array}{lr}
p = -{3 \over 2} n, & 2.7 \pm 0.0 \%, \\
{1 \over 2} \Delta^{++} -3n = 3 p, & 5.0 \pm 0.1 \%, \\
(p-n) = {1 \over 6} \Delta^{++} + 2 {1 \over \sqrt{2}} p \Delta^+, &
24.0 \pm 0.6  \% \\
\end{array}
\end{eqnarray}
which are obtained from Eq.~(\ref{su3coeffs}) by setting the coefficients $b_{(2)}^{1, {\bf 8}}$, $b_{(3)}^{1, {\bf 8}}$ and $c_{(3)}^{1, {\bf 8}}$ to zero.
These relations for strangeness zero baryons are expected to hold to relative order $1/N_c^2$, or approximately $10 \%$ accuracy, in the 
$SU(3)$ flavor symmetry limit~\cite{jl,djmtwo}.  

The additional magnetic moment relations
Eq.~(\ref{su6}) are predictions of contracted $SU(6)$ spin-flavor symmetry for large-$N_c$ baryons~\cite{jl,djmtwo}.  The relations are
identical to the $SU(6)$ magnetic moment relations of Beg, Lee and Pais~\cite{begleepais}, since only the single operator $G^{iQ}$ is retained in the $1/N_c$ expansion, which is the same operator as the non-relativistic quark model.
Contracted spin-flavor symmetry of large-$N_c$ baryons does not predict the absolute normalization of the baryon magnetic moments due to the arbitrary coefficient $a_{(1)}^{1, {\bf 8}}$ in Eq.~(\ref{1ops}).  The $1/N_c$ expansion gives a  
quantitative prediction for the accuracy of the spin-flavor relations in terms of factors of $1/N_c$ and $SU(3)$ flavor symmetry breaking.  
The $1/N_c$ expansion predicts that the $SU(6)$ relations of the strangeness zero baryons will be more accurate than $SU(6)$ relations between the magnetic moments of baryons with different strangeness, which have violations due to $SU(3)$ flavor symmetry breaking.  One example of such an $SU(6)$ relation~\cite{begleepais} is  
\begin{eqnarray}
&p = -\Omega^-, \quad
32.1 \pm 2.1 \% .&
\end{eqnarray}

\section{$SU(3)$ Breaking Analysis}

This section contains the
$SU(3)$ symmetry breaking operator analysis for the baryon magnetic moments.  The complete operator basis of the $1/N_c$ expansion contains 27 operators which parametrize the 27 baryon magnetic moments.  Symmetry-breaking operators 
are divided into operators which arise
at leading order in $SU(3)$ symmetry breaking, operators which occur at subleading orders in $SU(3)$ symmetry breaking, and
operators which violate isospin symmetry.

\subsection{Leading Order $SU(3)$ Breaking}

The baryon magnetic moment operator is a spin-1 flavor-octet operator in the $SU(3)$ flavor symmetry limit.  To leading order in $SU(3)$ breaking, all representations in
\begin{eqnarray}
(1,{\bf 8} \otimes {\bf 8}) = (1, {\bf 1} \oplus {\bf 8}_S \oplus {\bf 8}_A \oplus {\bf 10} \oplus \overline{\bf 10} \oplus {\bf 27} )
\end{eqnarray}
are obtained.  Each of these representations is considered separately.

\subsubsection{$(1,{\bf 8})$}

The four coefficients of the spin-1 flavor-${\bf 8}$ operators of the $SU(3)$ symmetry limit 
each receive $SU(3)$-violating contributions at one-loop in baryon chiral perturbation theory.
There are contributions which are order $m_q^{1/2}$ and order $m_q \ln m_q$ in the limit of degenerate decuplet and octet baryons, $\Delta M=0$.\footnote{For nonvanishing $\Delta M$, the functional dependence of each of these corrections generalizes to a known function ${\cal F}(m_q, \Delta M)$.  The chiral corrections are given by these more general  
functions of $m_q$ and $\Delta M$.}  
Because each of the four operators of the $1/N_c$ expansion transforms as the $a=Q$ component of a flavor octet, which is traceless, no flavor-singlet subtractions are required for these operators.

Four additional spin-1 flavor-${\bf 8}$ operators are obtained at leading order in $SU(3)$ breaking.  These operators are
\begin{eqnarray}
&&d^{Qb8} \left( \bar a^{1,{\bf 8}}_{(1)}\ G^{ib} + \bar b^{1,{\bf 8}}_{(2)}\ {1 \over N_c} J^i T^b 
+\bar b^{1,{\bf 8}}_{(3)}\  {1 \over N_c^2}{\cal D}_{(3)}^{ib} \right. \nn
&&\qquad\qquad \left. + \bar c^{1,{\bf 8}}_{(3)}\ {1 \over N_c^2}{\cal O}_{(3)}^{ib} \right),
\end{eqnarray} 
which reduce to
\begin{eqnarray}\label{8ops2}
&&{1 \over \sqrt{3}}  \left( \bar a^{1,{\bf 8}}_{(1)}\ G^{i\overline Q} + \bar b^{1,{\bf 8}}_{(2)}\ {1 \over N_c} J^i T^{\overline Q} 
+\bar b^{1,{\bf 8}}_{(3)}\  {1 \over N_c^2}{\cal D}_{(3)}^{i\overline Q} \right. \nn
&&\qquad\qquad \left. + \bar c^{1,{\bf 8}}_{(3)}\ {1 \over N_c^2}{\cal O}_{(3)}^{i\overline Q} \right),
\end{eqnarray} 
since
\begin{eqnarray}
d^{Qb8} O^{ib} &=& \left( d^{3b8}  + {1 \over \sqrt{3}} d^{8b8}\right) O^{ib} \nn
&=& {1 \over \sqrt{3}} \left( O^{i3} -{1 \over \sqrt{3}} O^{i8}\right)\equiv {1 \over \sqrt{3}} O^{i \overline Q},
\end{eqnarray}
where
\begin{eqnarray}
T^{\overline Q} &=& \pmatrix{ {1 \over 3} & 0 & 0 \cr
0 & -{2 \over 3} & 0 \cr
0 & 0 & {1 \over 3} \cr} = T^3 - {1 \over \sqrt{3}} T^8 \equiv \overline{Q}.
\end{eqnarray}
These $a=\overline Q$ flavor-octet operators also require no flavor-singlet subtractions since $\overline{Q}$ is traceless.  The 
operators arise in
chiral perturbation theory at one loop, with a nonanalytic quark mass dependence of $m_q^{1/2}$ for $\Delta M=0$.

\subsubsection{$(1,{\bf 10}+ {\overline {\bf 10}})$}

There are two spin-1 operators in the ${\bf 10}+ {\overline {\bf 10}}$ flavor representation,
\begin{eqnarray}\label{10ops}
c_{(2)}^{1,{\bf 10+ \overline{10}}} \ {1 \over N_c} O_{(2)}^{i[8Q]}
+ c_{(3)}^{1,{\bf 10+ \overline{10}}} \ {1 \over N_c^2} O_{(3)}^{i[8Q]},
\end{eqnarray}
where
\begin{eqnarray}
O_{(2)}^{i[8Q]} &\equiv& \left(\{T^8, G^{iQ} \}- \{ G^{i8}, T^Q \}\right) , \nn
O_{(3)}^{i[8Q]} &\equiv&  \left( \{ G^{i8}, J^k G^{kQ} \}- \{J^k G^{k8} ,G^{iQ}\}\right).
\end{eqnarray}
In general, the spin-1 flavor-$({\bf 10}+ {\overline {\bf 10}})$ operators
\begin{eqnarray}
O_{(2)}^{i[ab]} &\equiv& \left(\{T^a, G^{ib} \}- \{ G^{ia}, T^b \}\right) , \nn
O_{(3)}^{i[ab]} &\equiv&  \left( \{ G^{ia}, J^k G^{kb} \}- \{J^k G^{ka} ,G^{ib}\}\right),
\end{eqnarray}
require subtractions of the flavor-octet operators obtained by contracting with $f^{abc}$.   The specific operators
Eq.~(\ref{10ops}), however, do not require flavor-${\bf 8}$ subtractions since $f^{8Qc}=0$.  Because of the explicit antisymmetry of the flavor 
${\bf 10}+ {\overline {\bf 10}}$
operators in $[ab]$, only the $Q=3$ components of the operators in Eq.~(\ref{10ops}) are nonvanishing.  Consequently, no new operators arise when $Q \rightarrow \overline Q$ at higher order in $SU(3)$ breaking.
Each of the operators in Eq.~(\ref{10ops}) arises in chiral perturbation theory at one loop with a flavor dependence that is $m_q^{1/2}$
for $\Delta M =0$.

\subsubsection{$(1,{\bf 1})$}

There are two spin-1 flavor-singlet operators
\begin{eqnarray}\label{singletops}
\delta^{Q8} \left( c^{1,{\bf 1}}_{(1)}\ J^i +  c^{1,{\bf 1}}_{(3)}\ {1 \over N_c^2} \{J^2, J^i \} \right).
\end{eqnarray}
The leading contribution to these operator coefficients is a calculable chiral logarithm of order $m_q \ln m_q$ in the
limit that $\Delta M=0$.

\subsubsection{$(1,{\bf 27})$}

There are three spin-1 flavor-${\bf 27}$ operators generated at leading order in $SU(3)$ breaking,
\begin{eqnarray}\label{27ops1}
&&c_{(2)}^{1,{\bf 27}} \ {1 \over N_c} O_{(2)}^{i(8Q)}
+ b_{(3)}^{1,{\bf 27}} \ {1 \over N_c^2} D_{(3)}^{i(8Q)}\nn
&&\qquad + c_{(3)}^{1,{\bf 27}} \ {1 \over N_c^2} O_{(3)}^{i(8Q)},
\end{eqnarray}
where
\begin{eqnarray}\label{27opsdef}
O_{(2)}^{i(ab)} &\equiv& \left(\{T^a, G^{ib} \}+ \{ G^{ia}, T^b \}\right) , \nn
D_{(3)}^{i(ab)} &\equiv& J^i \{T^a, T^b \},\nn
O_{(3)}^{i(ab)} &\equiv&  \left( \{ G^{ia}, J^k G^{kb} \}+ \{J^k G^{ka} ,G^{ib}\}\right).
\end{eqnarray}
These flavor-${\bf 27}$ operators require flavor-octet and flavor-singlet subtractions.
All three operators arise in chiral perturbation theory at one loop from a calculable chiral logarithm of order $m_q \ln m_q$
in the limit $\Delta M = 0$.

\subsection{Subleading Order $SU(3)$ Breaking}

Additional operators are generated at subleading orders in $SU(3)$ breaking.  Most of the additional operators occur in the tensor product $(1,{\bf 8} \otimes {\bf 8} \otimes {\bf 8})$.  All of the additional operators transform as either $(1,{\bf 27})$ or $(1,{\bf 64})$ representations.

\subsubsection{$(1,{\bf 27})$}

There are three additional $(1, {\bf 27})$ operators which are of subleading order in $SU(3)$ breaking and which do not break isospin symmetry, 
\begin{eqnarray}\label{27ops2}
&&d^{Qb8} \left(\bar c_{(2)}^{1,{\bf 27}} \ {1 \over N_c} O_{(2)}^{i(8b)} + \bar b_{(3)}^{1,{\bf 27}} \ {1 \over N_c^2} D_{(3)}^{i(8b)} \right. \nn
&&\qquad\qquad \left. + \bar c_{(3)}^{1,{\bf 27}} \ {1 \over N_c^2} O_{(3)}^{i(8b)} \right) = \nn
&&{1 \over \sqrt{3}} \left(\bar c_{(2)}^{1,{\bf 27}} \ {1 \over N_c} O_{(2)}^{i(8 \overline Q)} 
+ \bar b_{(3)}^{1,{\bf 27}} \ {1 \over N_c^2} D_{(3)}^{i(8 \overline Q)} \right. \nn
&&\qquad\qquad \left. + \bar c_{(3)}^{1,{\bf 27}} \ {1 \over N_c^2} O_{(3)}^{i(8 \overline Q)} \right),
\end{eqnarray}
where the operators in this equation are defined in Eq.~(\ref{27opsdef}).
These operators first arise in chiral perturbation theory at two loops.  These flavor-${\bf 27}$ operators require flavor-${\bf 8}$ and flavor-${\bf 1}$ subtractions.

\subsubsection{$(1,{\bf 64})$}

There are two additional $(1, {\bf 64})$ operators at second subleading order in $SU(3)$ breaking which do not break isospin symmetry, 
\begin{eqnarray}\label{64ops1}
c_{(3)}^{1,{\bf 64}} \ {1 \over N_c^2} O_{(3)}^{i(88Q)} +d_{(3)}^{1,{\bf 64}}  {1 \over N_c^2} \tilde O_{(3)}^{i(88Q)},
\end{eqnarray}
where
\begin{eqnarray}
O_{(3)}^{i(88Q)} &\equiv& \{ G^{i8}, \{T^8, T^Q \}\}, \nn
\tilde O_{(3)}^{i(88Q)} &\equiv& \{T^8, \{ T^8, G^{iQ} \}\}.
\end{eqnarray}
For flavor ${\bf 64}$ operators, one needs to perform flavor ${\bf 1}$, ${\bf 8}$ and ${\bf 27}$ subtractions.

Additional spin-1 flavor-${\bf 64}$ operators arise at third subleading order via the tensor product $(1,{\bf 8} \otimes {\bf 8} \otimes {\bf 8} \otimes {\bf 8})$.  Naively, there are two additional operators, $d^{Qb8} \{ G^{i8}, \{T^8, T^{b} \} \} $ and $d^{Qb8} \{ T^8, \{T^8, G^{ib} \} \}$, which reduce to $1/\sqrt{3}$ times $\{ G^{i8}, \{T^8, T^{\overline Q} \} \}$ and $\{ T^8, \{T^8, G^{i\overline Q} \} \}$, respectively.
The $(1, {\bf 64})$ operators are not linearly independent, however, since they satisfy
the relation
\begin{eqnarray}
&&\{ G^{i8}, \{ T^8, T^Q \} \} - \{ G^{i8}, \{ T^8, T^{\overline Q} \} \} = \nn
&&\qquad  \{ T^8, \{ T^8, G^{iQ} \} \} - \{ T^8, \{ T^8, G^{i \overline Q} \} \}.
\end{eqnarray}
Consequently, there are only three independent $(1, {\bf 64})$ operators.
Without loss of generality, only the first of the two ${\overline Q}$ operators
\begin{eqnarray}\label{64ops3}
\bar c_{(3)}^{1,{\bf 64}}\  {1 \over \sqrt{3}}  \ {1 \over N_c^2}
O_{(3)}^{i(88 \overline Q)},
\end{eqnarray}
is included in the $1/N_c$ expansion, where
\begin{eqnarray}
O_{(3)}^{i(88 \overline Q)} =  \{ G^{i8}, \{T^8, T^{\overline Q} \} \} .
\end{eqnarray}

\subsection{Isospin Breaking}

Additional operators arise from isospin breaking, which transforms as the $a=3$ component of a flavor-${\bf 8}$ representation.
The only additional operators are either $(1,{\bf 27})$ or $(1,{\bf 64})$ operators.  There are no new $(1, {\bf 8})$ operators, since the operators
\begin{eqnarray}
d^{Qb3} \left( G^{ib} + {1 \over N_c} J^i T^b +{1 \over N_c^2}{\cal D}_{(3)}^{ib}
+ {1 \over N_c^2}{\cal O}_{(3)}^{ib} \right)
\end{eqnarray} 
reduce to
\begin{eqnarray}
d^{Qb3} O^{ib} &=& \left( d^{3b3}  + {1 \over \sqrt{3}} d^{8b3}\right) O^{ib} \nn
&=& {1 \over \sqrt{3}} \left( O^{i8} +{1 \over \sqrt{3}} O^{i3}\right),
\end{eqnarray}
which are linear combinations of the spin-1 flavor-${\bf 8}$ operators of Eqs.~(\ref{su3}) and~(\ref{8ops2}).

\subsubsection{$(1,{\bf 27})$}

The additional flavor-${\bf 27}$ operators which break isospin symmetry are 
\begin{eqnarray}\label{27ops3}
&&g_{(2)}^{1, {\bf 27}} {1 \over N_c} O_{(2)}^{i(3Q)} + f_{(3)}^{1, {\bf 27}} {1 \over N_c^2} D_{(3)}^{i(3Q)}\nn 
&&\qquad\qquad + g_{(3)}^{1, {\bf 27}} {1 \over N_c^2} O_{(3)}^{i(3Q)},
\end{eqnarray}
where
\begin{eqnarray}
O_{(2)}^{i(3Q)} &\equiv& \left( \{ T^3, G^{iQ} \} + \{ G^{i3}, T^Q \} \right), \nn
D_{(3)}^{i(3Q)} &\equiv& J^i \{ T^3, T^Q \}, \nn
O_{(3)}^{i(3Q)} &\equiv& \left( \{ G^{i3} , J^k G^{kQ} \} + \{ J^k G^{k3} , G^{iQ} \} \right).
\end{eqnarray}
These operators contain the $I=2$ operators $\{ T^3, G^{i3} \}$, $J^i \{ T^3, T^3 \}$, and
$\{ G^{i3} , J^k G^{k3} \}$.

\subsubsection{$(1,{\bf 64})$}

The additional flavor-${\bf 64}$ operators which break isospin symmetry are 
\begin{eqnarray}\label{64ops4}
&&g_{(3)}^{1, {\bf 64}}  {1 \over N_c^2} O_{(3)}^{i(33Q)} +  f_{(3)}^{1, {\bf 64}}  {1 \over N_c^2} \tilde O_{(3)}^{i(33Q)} \nn
&&\qquad + \bar g_{(3)}^{1, {\bf 64}} {1 \over N_c^2} O_{(3)}^{i(33 \overline Q)},
\end{eqnarray}
where
\begin{eqnarray}
O_{(3)}^{i(33Q)} &\equiv& \{ G^{i3}, \{ T^3, T^Q \} \}, \nn
\tilde O_{(3)}^{i(33Q)} &\equiv& \{ T^3, \{ T^3, G^{iQ} \} \}, \nn
O_{(3)}^{i(33\overline Q)} &\equiv& \{ G^{i3}, \{ T^3, T^{\overline Q} \} \}.
\end{eqnarray}
These operators contain the $I=2$ and $I=3$ operators
$\{ G^{i3}, \{ T^3, T^8 \} \}$, $\{ T^3, \{ T^3, G^{i8} \} \}$, and $\{ G^{i3}, \{ T^3, T^3 \} \}$.

\bigskip
In summary,
there are 27 independent operators in the $1/N_c$ expansion when isospin violation is included.
Four operators occur in the $SU(3)$ flavor symmetry limit, Eq.~(\ref{su3}).  The $SU(3)$ flavor-symmetry breaking operators consist of four $(1, {\bf 8})$ operators, two $(1, {\bf 10} + {\bf \overline{10}})$ operators, two $(1, {\bf 1})$ operators, 
six $(1, {\bf 27})$ operators, and three $(1, {\bf 64})$ operators.  These 17 operators appear in Eqs.~(\ref{8ops2}), (\ref{10ops}),
(\ref{singletops}), (\ref{27ops1}), (\ref{27ops2}), (\ref{64ops1})  and (\ref{64ops3}).  There are six additional operators which break isospin symmetry, Eqs.~(\ref{27ops3}) and~(\ref{64ops4}).  If one works in the isospin symmetry limit, there are only 21 independent baryon magnetic moments, and the six isospin-violating operators are not included in the $1/N_c$ expansion.  

The matrix elements of the 27 operators of the $1/N_c$ expansion for the baryon magnetic moments are given in Tables~I and~II.
Table~I contains the 15 operators which arise at leading order in $SU(3)$ breaking, whereas Table~II contains the remaining 12 operators.
The first six operators of Table~II violate $SU(3)$ symmetry at subleading orders, while the last six operators violate isospin symmetry.
The bosonic operator method of Ref.~\cite{bosonic} was useful for the calculation of the operator matrix elements.

The implications of this operator basis are studied in the next section.

\section{Results}

Each of the 27 operators of the $1/N_c$ expansion contributes to one linear combination of baryon magnetic moments.  These combinations are tabulated in Tables~III, IV, and~V.  For convenience, Tables~VI and~VII give the combinations for the nine $(1, {\bf 27})$ operators with isospin $I=0$, $1$, and $2$, and the six $(1, {\bf 64})$ operators with isospin $I=0$, $1$, $2$ and $3$, respectively.
Since many of the baryon magnetic moments are not measured, the majority of these linear combinations can not be evaluated at present.
Each of the 27 operators of the $1/N_c$ expansion contributes at a specific order in $1/N_c$ given by the explicit factor $1/N_c^{n-1}$ times
the $N_c$-dependence of its operator matrix elements $\langle {\cal O}_{(n)} \rangle$.  In addition, each operator of the $1/N_c$ expansion transforms as a specific representation of $SU(3)$ flavor symmetry, and so its coefficient $c_{(n)}$ is a function of the $SU(3)$ breaking parameters, the light quark masses $m_q$.  Most of the operators first arise from chiral loop corrections which are nonanalytic in the light quark masses, and their operator coefficients are proportional to these calculable nonanalytic functions.

The leading
$1/N_c$ and $m_q$ dependences of each operator and its operator coefficient are summarized in Tables~VIII and~IX.  Table~VIII contains the 21
operators which 
respect isospin symmetry, whereas Table~IX contains the six isospin-violating operators.  
The $SU(3)$ breaking and isospin breaking factors listed in Tables~VIII and~IX, respectively, give the functional dependences on $m_q$ of the symmetry-breaking factors, which is valid for $\Delta M=0$.   The actual functional dependence is more complicated than this, depending nonanalytically on both the light quark masses $m_q$ through the pion masses $M_\Pi$ and on the decuplet-octet singlet hyperfine mass splitting $\Delta M$.  Closed form expressions for the these generalized flavor symmetry breaking functions  for nonvanishing $\Delta M$ are known, and can be found in the literature.  Since these functions will not be numerically evaluated in this work, they are not given explicitly in the tables.  The $1/N_c$ dependences of the operators given in Tables~VIII and~IX follow from the fact that matrix elements of
$G^{i3}$ are order $N_c$, whereas matrix elements of $G^{i8}$ are order unity.  $T^3$ has matrix elements of order unity, and $T^8$ has matrix elements of order unity in their contribution to baryon magnetic moment splittings.  The identities
\begin{eqnarray}
\{ J^i, G^{i8} \} &=& {1 \over {2\sqrt{3}}} \{ J^i, J^i - 3 J_s^i \} \nn
&=& {1 \over {2\sqrt{3}}} \left( -J^2 - 3 J_s^2 + 3 I^2 \right),\nn
\{ J^i, G^{i3} \} &=& {1 \over 2} \left( V^2 - U^2 + J_u^2 - J_d^2 \right),
\end{eqnarray} 
where $V^2$ and $U^2$ refer to $V$-spin and $U$-spin, are useful for determining some of the operator matrix elements and their $N_c$ dependences.
The matrix elements of $\{ J^i, G^{i8} \}$ are order unity, since spin $J$, strange spin $J_s$ and isospin $I$ are all order unity for the large-$N_c$ baryons.  The matrix elements of $\{J^i, G^{i3} \}$ are order $N_c$ at leading order, since for large-$N_c$ baryon flavor representations, both $J_u^2$ and $J_d^2$
are order $N_c^2$, but their difference is order $N_c$.

From Tables~VIII and~IX, one is able to determine which operators are the leading operators in the combined expansion in $1/N_c$ and flavor symmetry breaking.  There are two operators which contribute at leading order $N_c$ to the baryon magnetic moments, namely $G^{iQ}$ and $G^{i \overline Q}$.  The $G^{iQ}$ respects $SU(3)$ flavor symmetry, and therefore is order unity in $SU(3)$ breaking.  In contrast, the operator $G^{i \overline Q}$ breaks $SU(3)$ flavor symmetry.
The leading $G^{i \overline Q}$ operator contribution arises from a one-loop chiral diagram Fig.~\ref{fig:1} which depends nonanalytically on the light quark masses as $m_q^{1/2}$.
Thus, the operator $G^{i \overline Q}$ is suppressed relative to the operator $G^{iQ}$ by $SU(3)$ breaking, but it is not suppressed in 
$1/N_c$.  It is important to emphasize that $SU(3)$-breaking also contributes to $G^{iQ}$ at one-loop order in chiral perturbation theory, so its coefficient receives subleading contributions proportional to $m_q^{1/2}$ and $m_q \ln m_q$ from one-loop pion graphs in addition to its leading $SU(3)$ symmetric contribution of order unity.  

The first 15 operators in Table~VIII arise at leading order in $SU(3)$ breaking
from one-loop pion corrections in $SU(3)$ chiral perturbation theory.  The largest $SU(3)$ breaking contribution is proportional to $m_q^{1/2}$, and the $\overline Q$ flavor-octet and the flavor-${\bf 10 + \overline {10}}$ operators receive leading contributions proportional to this $SU(3)$ breaking factor.  The flavor-singlet and 
three flavor-27 operators receive chiral logarithmic contributions proportional to $m_q \ln m_q$ at leading order in $SU(3)$ breaking.  The 15 operators which occur at leading order in $SU(3)$ breaking occur at various orders in $1/N_c$, as listed in Table~VIII.  
As mentioned already, there are two leading $O(N_c)$ operators, $G^{iQ}$ and $G^{i \overline Q}$.  The next largest operators in the 
$1/N_c$ hierarchy are the three
$O(1)$ operators, each of which occurs at leading order in $SU(3)$ breaking.  These operators are the $2$-body $(1, {\bf 10 + \overline {10}})$ operator, $J^i$ and the $2$-body $(1, {\bf 27})$ operator.  The $2$-body $(1, {\bf 10 + \overline{10}})$ operator coefficient is proportional to the $SU(3)$ breaking factor $m_q^{1/2}$; the two other $O(1)$ operators generated at leading order in $SU(3)$ breaking are proportional to $m_q \ln m_q$.  At order $1/N_c$ in the $1/N_c$ expansion, there are eight additional operators which are leading order in $SU(3)$ breaking.  The two remaining operators which are leading order in $SU(3)$ breaking are
$\{ J^2, J^i \}$ and $J^i \{ T^8, T^Q \}$, which are order $1/N_c^2$ in the $1/N_c$ expansion.  All of the other $SU(3)$-violating operators in Table~VIII first arise at subleading orders in $SU(3)$ chiral perturbation theory, so
the dominant $SU(3)$ breakings proportional to the $m_q^{1/2}$, $m_q \ln m_q$ and $m_q$ from one-loop chiral graphs and from counterterms with one insertion of the quark mass matrix
do not generate these subleading operators.  

Table~IX presents the corresponding $1/N_c$ and isospin-breaking factors for the six isospin-violating operators of the $1/N_c$ expansion.  Three of these operators
occur at one-loop in chiral perturbation theory, and have coefficients proportional to isospin-breaking with $m_q \ln m_q$ dependence.  
The remaining three operators arise at two-loop or three-loop order in chiral perturbation theory.  Since isospin breaking is much smaller than $SU(3)$ breaking and $1/N_c$, the six isospin-violating operators are much more suppressed than the 21 operators of Table~VIII.     
  
\subsection{Fits}

The $1/N_c$ and flavor breaking hierarchy given by Tables~VIII and~IX suggests a number of fits to the experimental data.
The first fit, Fit A, is the conventional four parameter $SU(3)$ symmetric fit, which is used for comparison with the subsequent
analysis. The $SU(3)$  flavor symmetry breaking parameter $m_q^{1/2}$ is roughly of order $1/\sqrt{N_c}$. This suggest a unified expansion in both flavor symmetry breaking and $1/N_c$,  treating $m_q \sim 1/\sqrt{N_c}$  and $m_q \ln m_q \sim 1/N_c$.  The successive fits, Fits B, C, D, E and F of Tables~X and~XI correspond to using this unified power counting. Fit B retains the $O(N_c)$ term; Fit C adds a $O(\sqrt{N_c})$ term; Fit D adds a $O(1/\sqrt{N_c})$ term; Fit E adds five $O(1/N_c)$ terms and Fit F adds four $O(1/N_c^{3/2})$ terms. There are three operators of leading order in $SU(3)$ breaking in Table~II that are left out of Fit F, one is 
$O(1/N_c^2)$ and two are $O(1/N_c^3)$ in the combined $1/N_c$ and $SU(3)$ breaking expansion, and so are higher order in the unified power counting. Only thirteen magnetic moments have been measured.  Once two more magnetic moments are measured, a determination of all of the 15 leading $SU(3)$ breaking operators of Table~II, including the two higher order $SU(3)$ symmetric operators not included in Fit F,  becomes possible.  

Fits A, B, C, D, E and F are fits to the 13 experimental data points listed in the first column of Table~X.  Since many of the magnetic moments used in the fits are measured much more accurately than the expected theoretical uncertainties, a theory uncertainty of $\sigma^{\rm theory}$ is added in quadrature to each experimental measurement to avoid biasing the fits.  The theory error added for each fit is tuned to yield $\chi^2/{\rm dof} = 1$ for the fit.  Consequently, $\sigma^{\rm theory}$ is a measure of how well the data fits the theory formula.  The primed fits, Fits ${\rm A}^\prime$, ${\rm B}^\prime$, ${\rm C}^\prime$, ${\rm D}^\prime$ and ${\rm E}^\prime$, are fits to the 10 most accurately measured magnetic moments.  These fits exclude the three primed data points in the first column of Table~X, which are the less well-measured $\Delta^{++}$ magnetic moment and the CLAS measurements of $\Lambda \Sigma^{*0}$ and $\Sigma \Sigma^{*+}$.   Also included in Table~X are the numerical values of all 27 magnetic moments evaluated at the central values of the extracted coefficients. 

\subsubsection{Fit A}

The first fit, Fit A, 
is the four-parameter $SU(3)$ symmetric fit using Eq.~(\ref{su3}).  Fit A yields the coefficients 
\begin{eqnarray}
a_{(1)}^{1, {\bf 8}} &=& 4.98 \pm 0.49, \nn
b_{(2)}^{1, {\bf 8}} &=& 0.66 \pm 1.49, \nn
b_{(3)}^{1, {\bf 8}} &=& -0.25 \pm 0.95, \nn
c_{(3)}^{1, {\bf 8}} &=& 4.18 \pm 1.71,
\end{eqnarray} 
with significant error bars, and a theory error of
$\sigma^{\rm theory} = 0.35\ \mu_N$.
Fit ${\rm A}^\prime$ is a fit to only the 10 of the 13 measured magnetic moments.  Values for the coefficients consistent with those of Fit A are obtained, and
$\sigma^{\rm theory} = 0.28 \ \mu_N$.  
The coefficient values obtained for Fits A and ${\rm A}^\prime$ also are consistent with the earlier determination using strangeness zero baryon magnetic moments Eq.~(\ref{abbccoeffs}), within their rather large error bars.  Again, it is important to note that the $\Delta^{++}$ magnetic moment is not included in Fit ${\rm A}^\prime$, whereas the determination of Eq.~(\ref{abbccoeffs}) depends upon the 
$\Delta^{++}$ magnetic moment (see Eq.~(\ref{su3coeffs})).
There is tension between the $\Delta^{++}$ and $\Omega^-$ measurements and between the 
$p \Delta^+$ and the transition magnetic moments $\Lambda \Sigma^{*0}$ and $\Sigma \Sigma^{*+}$ in an $SU(3)$ symmetry fit,
since $SU(3)$ relations imply that these magnetic moments are simply related.  The
Fit ${\rm A}^\prime$, which does not contain this additional data, therefore finds smaller $SU(3)$ breaking, but this result is not physically significant.  Significant $SU(3)$ breaking in the decuplet and the transition magnetic moments is allowed.   

\subsubsection{Fit B}

Fit B is a one-parameter fit of the data to the single operator 
$G^{iQ}$, which is the leading operator of the $SU(3)$ symmetric limit and of the combined $1/N_c$ and flavor symmetry breaking expansion.  
From Fit B, one extracts the leading operator coefficient
\begin{eqnarray}\label{aone}
a_{(1)}^{1, {\bf 8}} = 5.45 \pm 0.31 ,
\end{eqnarray} 
with $\sigma^{\rm theory}= 0.41 \ \mu_N$.  For the one-parameter fit, again there is tension between obtaining the experimental values of both $\Delta^{++}$ and $\Omega^-$ magnetic moments in an $SU(3)$ symmetric fit, and there is tension between obtaining $p \Delta^+$ and the new two CLAS measurements.
The extracted value of $a_{(1)}^{1, {\bf 8}}$ of Fit B is consistent with the experimental determination of $(5.88 \pm 0.09)$ from the strangeness zero baryon magnetic moments of Eq.~(\ref{su3coeffs}) in Section~I.  Fit ${\rm B}^\prime$ including only 10 of the 13 magnetic moments yields the value
 \begin{eqnarray}
a_{(1)}^{1, {\bf 8}} = 5.07 \pm 0.35 ,
\end{eqnarray} 
with $\sigma^{\rm theory} = 0.39 \ \mu_N$.  
The greater discrepancy in this extracted value with the determination of Eq.~(\ref{su3coeffs}) occurs because the latter determination depends upon the $\Delta^{++}$ magnetic moment, which is not one of the 10 magnetic moments included in Fit ${\rm B}^\prime$.

\subsubsection{Fit C}

Fit C is a two-parameter fit to the leading operator $G^{iQ}$ and the first subleading operator $G^{i \overline Q}$.  Both of these operators are $O(N_c)$.  The $G^{iQ}$ operator respects $SU(3)$ flavor symmetry.  The $G^{i \overline Q}$ operator violates $SU(3)$ symmetry at leading order $m_q^{1/2}$.  A theory error of $0.27 \ \mu_N$ is found for Fit C, so Fit C is a better fit to the data than the
four-parameter $SU(3)$ symmetry fit, Fit A.
The extracted coefficients are
\begin{eqnarray}
a_{(1)}^{1, {\bf 8}} &=& 5.07 \pm 0.23, \nn
\bar a_{(1)}^{1, {\bf 8}} &=& 1.60 \pm 0.42 \ .
\end{eqnarray} 
The suppression of the $\bar a_{(1)}^{1, {\bf 8}}$ coefficient relative to the $a_{(1)}^{1, {\bf 8}}$ coefficient by a factor of $0.3$ is a reflection of the $SU(3)$-breaking suppression factor $m_q^{1/2}$, since both operators are $O(N_c)$.  Fit ${\rm C}^\prime$ to only 10 of the 13 magnetic moments yields similar values for the two coefficients,
\begin{eqnarray}
a_{(1)}^{1, {\bf 8}} &=& 4.85 \pm 0.21, \nn
\bar a_{(1)}^{1, {\bf 8}} &=& 1.54 \pm 0.37 ,
\end{eqnarray} 
and for the theory error, $\sigma^{\rm theory}= 0.23 \ \mu_N$.
These 2-parameter fits provide clear evidence for the $1/N_c$ and $SU(3)$ flavor breaking hierarchy of the combined $1/N_c$ and 
$SU(3)$ breaking expansion.  The unified expansion predicts that the  
dominant $SU(3)$ breaking in the magnetic moments is due to the $G^{i\overline Q}$ operator, and Fits C and ${\rm C}^\prime$ support this prediction.  For this 2-parameter fit, all baryon magnetic moments except $\Sigma^{*0}$ are nonzero.

\subsubsection{Fit D}

Fit D adds one additional operator to the fit, namely the $(1, {\bf 10 + \overline {10}})$ $2$-body operator, which is $O(1)$ in $1/N_c$ and
proportional to $m_q^{1/2}$.  All of the other operators proportional to $m_q^{1/2}$ flavor symmetry breaking which are not included in the fit are $O(1/N_c)$, and therefore suppressed by one factor of $1/N_c$ relative to the operators included in Fit D.  
Fit D has a theory error of $0.275 \ \mu_N$, and so it also is a better fit to the data than Fit A, even though it has fewer parameters.  Fit D yields coefficients 
$a_{(1)}^{1, {\bf 8}}$ and $\bar a_{(1)}^{1, {\bf 8}}$ essentially identical to those of Fit C, with identical error bars.  The third coefficient 
$c_{(2)}^{1, {\bf 10 + \overline{10}}}$ is determined to be $0.37 \pm 0.52$ in Fit D and $0.71 \pm 0.40$ in Fit ${\rm D}^\prime$.
The three operators of Fit D do not contribute to the $\Sigma^{*0}$ magnetic moment, which remains zero.

\subsubsection{Fit E}

Fit E is an eight-parameter fit which adds to the three operators of Fit D, the three $1/N_c$ suppressed operators of the $SU(3)$ symmetric fit Eq.~(\ref{su3}), $J^i$ and the leading $2$-body $(1, {\bf 27})$ operator.  This eight-operator fit includes
all of the $SU(3)$ breaking contributions occurring at one loop in chiral perturbation theory which are leading order in $1/N_c$.  Relative to Fit D, the five
added operators are the three $O(1/N_c)$ $SU(3)$-symmetric $(1, {\bf 8})$ operators Eq.~(\ref{su3}) which are $O(1)$ in $SU(3)$ flavor symmetry breaking, the $(1, {\bf 1})$ operator
$J^i$ which is $O(1)$ in $1/N_c$ and proportional to $m_q \ln m_q$, and the leading $2$-body $(1, {\bf 27})$ operator which is $O(1)$ in $1/N_c$ and proportional to $m_q \ln m_q$.  
The theory error $\sigma^{\rm theory}$ drops dramatically for Fit E to $0.024 \ \mu_N$.

For Fit ${\rm E}^\prime$, the theory error is $0.035 \ \mu_N$, so Fit E is a better than Fit ${\rm E}^\prime$.  One can view this result as evidence that $SU(3)$ breaking in the decuplet and decuplet-octet transition magnetic moments is substantial, since the primed fit leaves out the data which indicates significant $SU(3)$ breaking for the decuplet and transition magnetic moments.  For Fits E and ${\rm E}^\prime$,  the values for the coefficients already obtained in the earlier fits are mainly stable, and the experimental data is reproduced
at the central values of the coefficients to within the theory errors for the included data points.  
The $\Sigma^{*0}$ is nonvanishing for Fit E, due to the inclusion of the $J^i$ operator.  From Table~II matrix elements of the 15 leading operators in $SU(3)$ breaking, one sees that only the $J^i$ operator of these 15 leading operators contributes to the $\Sigma^{*0}$ magnetic moment.  Thus, this magnetic moment is a sensitive probe of the $J^i$ operator.  

\subsubsection{Fit F}

Fit F corresponds to a 12 operator fit.  Relative to Fit E, it adds four operators, namely the three $O(1/N_c)$ flavor-${\bf 8}$ operators in the $\overline Q$ flavor direction, proportional to $m_q^{1/2}$, as well as the $3$-body $(1, {\bf 10 + \overline{10}})$ operator which also is $O(1/N_c)$ and proportional to $m_q^{1/2}$.  Fit F necessarily includes all 13 measured magnetic moments.  A theory error of $0.06 \ \mu_N$ is found for the fit, so Fit F is not a better fit to the data than Fit E.  The coefficients of the operators which are present in Fit E are stable, with the exception of $c_{(3)}^{1, {\bf 8}}$.  The coefficients of the four additional operators in Fit F, however, have very significant error bars, and are not well determined.  The large errors on $c_{(3)}^{1, {\bf 8}}$, $\bar c_{(3)}^{1, {\bf 8}}$, and $c_{(3)}^{1, {\bf 10 + \overline{10}}}$ reflects degeneracies in fitting the decuplet-octet transition magnetic moments.    Only three of these transition magnetic moments are measured, and the values of these three ill-determined coefficients of Fit F can compensate for one another to return the three experimental values.  Measurement of additional transition magnetic moments in the future are necessary to break these degeneracies.     

\bigskip

One clear result of the fits is that the experimental data is reproduced by the operators of leading order in the combined expansion in 
$1/N_c$ and $SU(3)$ breaking.  There is evidence for both the $1/N_c$ and $SU(3)$ breaking pattern in the experimental data.
Fits to the unified expansion with fewer parameters, Fits C and D, are better fits to the data than the $SU(3)$ symmetry fit, Fit A.  
The extracted values of the coefficients of the combined expansion are stable when more operators are added, with the exception of Fit F, which suffers from degeneracies.  The eight operator fit, Fit E, is able to reproduce the experimental data within a theory error of $0.025 \ \mu_N$.

\subsection{$SU(3)$ Hierarchy}

Finally, it is worth returning to the magnetic moment linear combinations of Tables~~III, IV and~V.  It is not possible to evaluate most of the linear combinations because of their dependence on unmeasured magnetic moments.  However, it is possible to consider the various linear combinations of the octet baryon magnetic moments in an expansion in $SU(3)$ breaking alone.  These combinations yield a hierarchy of relations in $SU(3)$ breaking for the octet baryon magnetic moments.
The $SU(3)$ breaking analysis of the decuplet magnetic moment combinations and of the decuplet-octet transition magnetic moment combinations gives analogous relations for these magnetic moments.

First, consider the baryon octet magnetic moment combinations in an expansion in $SU(3)$ breaking.
At order $m_q^{1/2}$ in the $SU(3)$ chiral expansion, three $SU(3)$ combinations vanish, namely
\begin{eqnarray}
&2N_0 + \Lambda_0 + 3 \Sigma_0 + 2 \Xi_0,&\nn
&2N_0 - 3 \Lambda_0 - \Sigma_0 + 2 \Xi_0,& \nn
&N_1 + 2 \sqrt{3} \left( \Lambda \Sigma \right)_1 - \Xi_1.&
\end{eqnarray}
These three magnetic moment relations were first derived by Caldi and Pagels~\cite{Caldi} using $SU(3)$ chiral perturbation theory
for the octet baryons.  In Ref.~\cite{JLMS}, it was shown that these relations continue to hold when the decuplet baryons are included in heavy baryon chiral perturbation theory~\cite{jmhblone,jmhbltwo}. 
Including order $m_q \ln m_q$ and order $m_q$ counterterm effects, a single combination of the octet baryon magnetic moments continues to vanish.
This combination,
\begin{eqnarray}
\left[ N_1 + 2 \sqrt{3}  \left( \Lambda \Sigma \right)_1 - \Xi_1 \right] - \left[ 2 N_0 -3\Lambda_0 -\Sigma_0 + 2 \Xi_0 \right],
\end{eqnarray}
is equal to
\begin{eqnarray}\label{specialoctetreln}
-2n + 3 \Lambda + 2 \sqrt{3} \Lambda\Sigma^0 +{1 \over 2} \left( \Sigma^+ + \Sigma^- \right) -2 \Xi^0,
\end{eqnarray}
which is the same magnetic moment combination first found in Ref.~\cite{JLMS}.

Repeating this same $SU(3)$ symmetry breaking analysis for the baryon decuplet magnetic moments at order $m_q^{1/2}$ in the $SU(3)$ chiral expansion yields five linear combinations which vanish,
\begin{eqnarray}
&4 \Delta_0 + 3 \Sigma^*_0 + 2 \Xi^*_0 + \Omega_0,&\nn
&\Delta_1 - 3 \Sigma^*_1 - 4 \Xi^*_1,&\nn
&4 \Delta_0 -5 \Sigma^*_0 -2 \Xi^*_0 + 3 \Omega_0,&\nn
&\Delta_0 - 3 \Sigma^*_0 + 3 \Xi^*_0 - \Omega_0,&\nn
&\Delta_1 - 10 \Sigma^*_1 + 10 \Xi_1^*.&
\end{eqnarray}
Including order $m_q \ln m_q$ and order $m_q$ effects, three of the combinations
\begin{eqnarray}
&\left[ \Delta_1 - 3 \Sigma_1^* - 4 \Xi_1^* \right] -2 \left[ 4 \Delta_0 - 5 \Sigma_0^* - 2 \Xi_0^* + 3 \Omega_0 \right],&\nn
&\Delta_0 - 3 \Sigma^*_0 + 3 \Xi^*_0 - \Omega_0,&\nn
&\Delta_1 - 10 \Sigma^*_1 + 10 \Xi_1^*,&
\end{eqnarray}  
continue to vanish.

A similar $SU(3)$ analysis applies to the decuplet-octet transition magnetic moments.
At order $m_q^{1/2}$ in the $SU(3)$ chiral expansion, the three linear combinations
\begin{eqnarray}
& \left( \Sigma \Sigma^* \right)_0 - \left( \Xi \Xi^* \right)_0,&\nn
& \left( N \Delta \right)_1 -2 \sqrt{3} \left( \Lambda \Sigma^* \right)_1 + 3 \left( \Sigma \Sigma^* \right)_1 - 2 \left( \Xi \Xi^* \right)_1,&\nn
&-\left( N \Delta \right)_1 + 2 \sqrt{3} \left( \Lambda \Sigma^* \right)_1 + \left( \Sigma \Sigma^* \right)_1 -2 \left( \Xi \Xi^* \right)_1,&
\end{eqnarray}
vanish.  Including order $m_q \ln m_q$ and order $m_q$ effects, two of these combinations,  
\begin{eqnarray}
&&\left[ \left( N \Delta \right)_1 -2 \sqrt{3} \left( \Lambda \Sigma^* \right)_1 + 3 \left( \Sigma \Sigma^* \right)_1 - 2 \left( \Xi \Xi^* \right)_1 \right] \nn
&&\qquad\qquad -8\left[\left( \Sigma \Sigma^* \right)_0 - \left( \Xi \Xi^* \right)_0\right],\nn
&& -\left( N \Delta \right)_1 +2 \sqrt{3} \left( \Lambda \Sigma^* \right)_1 + \left( \Sigma \Sigma^* \right)_1 - 2 \left( \Xi \Xi^* \right)_1,
\end{eqnarray}
continue to vanish.

This $SU(3)$ hierarchy of the baryon magnetic moments represents a subhierarchy of the combined $1/N_c$ and $SU(3)$ breaking expansion.  Numerical evaluation of the baryon octet magnetic moment combinations and their experimental accuracies supports the conclusion of this work that the $SU(3)$ hierarchy alone is not sufficient to understand the spin-flavor structure of the baryon magnetic moments.  For example, the isovector combination 
\begin{eqnarray}
N_1 -\Sigma_1 + \Xi_1
\end{eqnarray}
in the $(1, {\bf 10 + \overline{10}})$ representation has an experimental accuracy of $11.0 \pm 0.7 \%$, which is to be compared with $1/N_c$ times the $SU(3)$ suppression factor corresponding to $m_q^{1/2}$ breaking.  Another example is the isovector combination in the $(1,{\bf 8})$ representation,
\begin{eqnarray}
-N_1 + 2 \sqrt{3} \left( \Lambda \Sigma \right)_1 + 4 \Sigma_1 + 5 \Xi_1,
\end{eqnarray}
which has an experimental accuracy of $8.6 \pm 2.2 \%$.  This combination is order unity in $SU(3)$ breaking, but is suppressed by 
$1/N_c^2$ relative to the leading $O(N_c)$ contribution to the isovector magnetic moments.  For the isoscalar combinations, the combination
\begin{eqnarray}
7 N_0 + 2 \Lambda_0 - 6 \Sigma_0 - 3 \Xi_0,
\end{eqnarray}   
is order unity in $SU(3)$ breaking, but suppressed by $1/N_c$ relative to the leading $O(1)$ contribution to the isoscalar magnetic moments.  This combination has an experimental accuracy of $14.7 \pm 1.5 \%$.  In each of these cases, the experimental accuracy
of the relation is not explained by $SU(3)$ breaking alone.

\section{Conclusions}

There are several clear results of the analysis of the baryon magnetic moments in a combined expansion in $1/N_c$ and $SU(3)$ flavor symmetry breaking of this work.  First, there is clear evidence in the experimental data for the combined hierarchy from the fits of Tables~X and~XI.
The successive fits of the combined expansion are Fits B, C, D, and E.  There is not enough experimental data to resolve the rest of the hierarchy at present.  Fit E is the most reliable fit obtained in this work.  
Second, the dominant $SU(3)$ flavor symmetry breaking for the baryon magnetic moments is due to the leading   
$O(N_c m_q^{1/2})$ contribution of $SU(3)$ chiral perturbation theory.  This $SU(3)$ breaking is described by the operator 
$G^{i \overline Q}$.  The next largest $SU(3)$ breaking is a $m_q^{1/2}$ breaking at $O(1)$ in the $1/N_c$ expansion from the
the leading $2$-body $(1, {\bf 10 + \overline{10}})$ operator $O_{(2)}^{i[8Q]}$.  

The nonanalytic $m_q^{1/2}$ chiral correction to the baryon magnetic moments is similar to the $O(m_q^{3/2})$ correction to the baryon masses, in that it is not suppressed in $1/N_c$.  In the case of the magnetic moments, however, the $m_q^{1/2}$ contribution is the leading order term in $SU(3)$ breaking, whereas the $m_q^{3/2}$ correction for the baryon masses is subdominant relative to the analytic $O(m_q)$ contribution.  This fact makes the baryon magnetic moments a particularly useful observable for studying $SU(3)$ flavor symmetry breaking, since $SU(3)$ breaking is enhanced.  

The analysis of this work is a group theoretic one.  A hierarchy of baryon magnetic moments is obtained in terms of linear combinations of the 27 baryon magnetic moments.  The complete set of magnetic moment combinations to all orders in $1/N_c$ and $SU(3)$ flavor symmetry breaking is obtained.  
Further progress now requires additional experimental measurements.  
The very precisely measured baryon octet magnetic moments contain substantial $SU(3)$ flavor symmetry breaking.
The measurements of the decuplet baryon magnetic moments $\Omega^-$ and $\Delta^{++}$ also indicate significant $SU(3)$ breaking, although the error bar on the $\Delta^{++}$ magnetic moment is still quite large.    
The recent CLAS measurements of the $\Lambda \Sigma^{*0}$ and $\Sigma \Sigma^{*+}$ transition magnetic moments reveal that there is substantial $SU(3)$ breaking in the transition magnetic moments as well.  
Fit E supports the conclusion that $SU(3)$ breaking of the decuplet and the decuplet-octet transition magnetic moments is substantial.
The $1/N_c$ expansion implies that certain linear combinations of the octet, decuplet and transition magnetic moments will be highly suppressed in $1/N_c$ and flavor symmetry breaking.  Further data is needed to verify these predictions.

The eight operator fit, Fit E, of Table XI is able to reproduce the experimental data to within a theory error of $0.025\ \mu_N$.  Estimates for all of the unmeasured baryon magnetic moments are obtained from this fit.  A number of features can be seen.  There is an equal space rule for the four $\Delta$ baryon magnetic moments implied by $SU(3)$ symmetry.  This equal spacing approximately persists even in the presence of $SU(3)$ breaking.  The $\Sigma^{*0}$ magnetic moment is particularly small.  It vanishes in the $SU(3)$ limit, and only receives a nonvanishing contribution from one of the fifteen operators of Table~I which is leading order in $SU(3)$ breaking.
The $\Sigma \Sigma^{*-}$ and the $\Xi \Xi^{*-}$ transition magnetic moments, which also vanish in the $SU(3)$ symmetry limit, also remain quite small even in the presence of $SU(3)$ breaking.

\acknowledgements

The author thanks Ken Hicks for bringing the recent CLAS measurements to her attention.  This work was supported in part by DOE grant DE-FG02-90ER40546.

\vfill\break\eject

\setlength{\arraycolsep}{2pt}
\renewcommand{\arraystretch}{1.7}

\begin{table*}
\caption{Matrix Elements of Magnetic Moment Operators for Exact $SU(3)$ Flavor Symmetry and to Leading Order in $SU(3)$
Flavor Symmetry Breaking.  In the $SU(3)$ symmetry limit, there are four operators $G^{iQ}$, $J^i T^Q$, 
${\cal D}_{(3)}^{i Q}$ and ${\cal O}_{(3)}^{i Q}$.  Leading order $SU(3)$ symmetry breaking contributes to all fifteen operators.}
\smallskip
\label{tab:15ops}
\begin{eqnarray*}
\begin{array}{|c|c|c|c|c||c|c|c|c|c|c|c|c|c|c|c|}
\hline
& G^{iQ} & J^i T^Q & {\cal D}_{(3)}^{i Q}  & {\cal O}_{(3)}^{i Q}
& J^i & \{J^2, J^i \} 
& G^{i\overline Q} & J^i T^{\overline Q} & {\cal D}_{(3)}^{i \overline Q}  & {\cal O}_{(3)}^{i \overline Q}
& {1 \over \sqrt{3}} O_{(2)}^{i[8Q]} & {1 \over \sqrt{3}} O_{(3)}^{i[8Q]} 
& {1 \over \sqrt{3}} O_{(2)}^{i(8Q)} & {1 \over \sqrt{3}} D_{(3)}^{i(8Q)} & {1 \over \sqrt{3}} O_{(3)}^{i(8Q)}
\\[2pt]
\hline
& (1, {\bf 8}) & (1, {\bf 8}) & (1, {\bf 8}) & (1, {\bf 8}) & (1, {\bf 1}) & (1, {\bf 1}) & (1, {\bf 8}) & (1, {\bf 8}) & (1, {\bf 8}) & (1, {\bf 8}) 
& (1, {\bf 10 + \overline{10}}) & (1, {\bf 10 + \overline{10}}) & (1, {\bf 27}) & (1, {\bf 27}) & (1, {\bf 27}) \\
\hline
 p
& {1 \over 2} & {1 \over 2} & {3 \over 2} & 0 
& {1 \over 2} & {3 \over 4} 
& {1 \over 3} & 0  & 1 & 0 
& {1 \over {3}}  & 0  
& {2 \over 3} & {1 \over 2}  & {1 \over 4}  
 \\
 n
& -{1 \over 3} & 0 & -1 & 0 
& {1 \over 2}  & {3 \over 4} 
& -{1 \over 2}   &  -{1 \over 2}   &  -{3 \over 2} & 0 
& -{1 \over {3}}  & 0  
& -{1 \over 3} & 0  & -{1 \over {6}} 
\\
 \Lambda
& -{1 \over 6} & 0 & -{1 \over 2} & 0 
& {1 \over 2}  & {3 \over 4} 
& {1 \over 6}  &  0 & {1 \over 2} & 0 
& 0  &  0 
& 0 & 0  & {1 \over {6}} 
\\
 \Lambda \Sigma^0
& -{1 \over {2\sqrt{3}}} & 0 & -{\sqrt{3} \over 2} & 0 
& 0 & 0 
& -{1 \over {2 \sqrt{3}}}  & 0  & -{\sqrt{3} \over 2} & 0 
& 0  &  0 
& 0 & 0 & 0 
\\
 \Sigma^+
& {1 \over 2} & {1 \over 2} & {3 \over 2} & 0 
& {1 \over 2}  & {3 \over 4} 
& {1 \over 6}  &  {1 \over 2}   & {1 \over 2} & 0  
& -{1 \over {3}}   & 0  
& {1 \over 3} & 0 & {1 \over 2} 
\\
 \Sigma^0
& {1 \over 6} & 0 & {1 \over 2} & 0 
& {1 \over 2}  & {3 \over 4} 
&  -{1 \over 6} &  0   & -{1 \over 2} &  0 
& 0  &  0 
& 0  &  0 & {1 \over 6} 
\\
 \Sigma^-
& -{1 \over 6} & -{1 \over 2} & -{1 \over 2} & 0 
& {1 \over 2}  & {3 \over 4} 
& -{1 \over 2}  &  -{1 \over 2}  & -{3 \over 2} & 0 
& {1 \over {3}}   &  0 
& -{1 \over 3} & 0 & -{1 \over {6}} 
\\
 \Xi^0
& -{1 \over 3} & 0 & -1 & 0  
& {1 \over 2}  & {3 \over 4} 
& {1 \over 6}  &  {1 \over 2}   & {1 \over 2} & 0  
&  {1 \over {3}}  & 0  
& {1 \over 3} & 0  & {1 \over 2} 
\\
 \Xi^-
& -{1 \over 6} & -{1 \over 2} & -{1 \over 2} & 0 
& {1 \over 2}  & {3 \over 4} 
& {1 \over 3} & 0  & 1 & 0  
&  -{1 \over {3}}  & 0  
& {2 \over 3} & {1 \over 2} & {1 \over 4} 
\\
 \Delta^{++}
& 1 & 3 & 15 & 0 
& {3 \over 2}  & {{45} \over 4} 
& {1 \over 2}  &  {3 \over 2}  & {{15} \over 2} & 0  
& 0  &  0 
& 2 & 3 & {{5} \over 2} 
\\
 \Delta^{+}
& {1 \over 2} & {3 \over 2}  & {{15} \over 2} & 0 
& {3 \over 2}  & {{45} \over 4} 
& 0  &  0  & 0 &  0 
& 0  &  0 
& 1 & {3 \over 2} & {5 \over 4} 
\\
 \Delta^{0}
& 0 & 0 & 0 & 0 
& {3 \over 2}  & {{45} \over 4} 
& -{1 \over 2}  &  -{3 \over 2}  & -{{15} \over 2} & 0  
& 0  &  0 
& 0 &  0  & 0 
\\
 \Delta^{-}
& -{1 \over 2} & -{3 \over 2} & -{{15} \over 2}  & 0 
& {3 \over 2}   & {{45} \over 4}  
& {-1} &  -{3} & -{15} & 0 
& 0  &  0 
& -1 & -{3 \over 2} & -{{5} \over 4} 
\\
 \Sigma^{*+}
& {1 \over 2} & {3 \over 2} & {{15} \over 2} & 0 
& {3 \over 2}   & {{45} \over 4}  
& {1 \over 2}  &  {3 \over 2}  & {{15} \over 2} & 0  
& 0  &  0 
& 0 &  0  & 0 
\\
 \Sigma^{*0}
& 0 & 0 & 0 & 0 
& {3 \over 2}   & {{45} \over 4}  
& 0  &  0  & 0 & 0  
& 0  &  0 
& 0 &  0  & 0 
\\
 \Sigma^{*-}
& -{1 \over 2} & -{3 \over 2}  & -{{15} \over 2} & 0 
& {3 \over 2}   & {{45} \over 4}  
& -{1 \over 2}  &  -{3 \over 2}   & -{{15} \over 2} & 0 
& 0  &  0 
& 0 &  0  & 0 
\\
 \Xi^{*0}
& 0 & 0 & 0 & 0 
& {3 \over 2}   & {{45} \over 4}  
& {1 \over 2}  &  {3 \over 2} & {{15} \over 2} & 0 
& 0  &  0 
& 0 &  0  & 0 
\\
 \Xi^{*-}
& -{1 \over 2} & -{3 \over 2}  & -{{15} \over 2} & 0 
& {3 \over 2}   & {{45} \over 4}  
& 0 &  0   & 0 & 0 
& 0  &  0 
& 1 & {3 \over 2} & {{5} \over 4} 
\\
 \Omega^-
& -{1 \over 2} & -{3 \over 2} & -{{15} \over 2} & 0 
& {3 \over 2}   & {{45} \over 4}  
& {1 \over 2}  &  {3 \over 2}  & {{15} \over 2} & 0 
& 0  &  0 
& 2 & {3}  & {{5} \over 2} 
\\
 {1 \over \sqrt{2}} p\Delta^+
& {1 \over 3} & 0 & 0 & {3 \over 2} 
& 0 & 0 
& {1 \over 3}  & 0  & 0 & {3 \over 2}  
& {1 \over {3}}  & -{1 \over 4}  
& {1 \over 3} & 0  & {1 \over 4} 
\\
 {1 \over \sqrt{2}} n\Delta^0
& {1 \over 3} & 0 & 0 & {3 \over 2} 
& 0 & 0 
& {1 \over 3}  &  0 & 0 & {3 \over 2}  
& {1 \over {3}}  & -{1 \over 4}  
& {1 \over 3} & 0  & {1 \over 4} 
\\
 {1 \over \sqrt{2}} \Lambda \Sigma^{*0}
& {1 \over {2\sqrt{3}}} & 0 & 0 & {{3\sqrt{3}} \over 4} 
& 0 & 0 
& {1 \over {2 \sqrt{3}}} &  0 & 0 & {{3\sqrt{3}} \over 4} 
& 0  &  0 
& 0 & 0 & -{1 \over {4\sqrt{3}}}  
\\
 {1 \over \sqrt{2}} \Sigma\Sigma^{*+}
& {1 \over 3} & 0 & 0 & {3 \over 2} 
& 0 & 0 
& 0 & 0  & 0 & 0 
& -{1 \over {3}}  & {1 \over 4}  
& {1 \over 3} & 0 & {5 \over {12}} 
\\
 {1 \over \sqrt{2}} \Sigma\Sigma^{*0}
& {1 \over 6} & 0 & 0 & {3 \over 4} 
& 0 & 0 
& -{1 \over 6} & 0  & 0 & -{3 \over 4} 
& 0  &  0 
& 0 & 0 & {1 \over {12}} 
\\
 {1 \over \sqrt{2}} \Sigma\Sigma^{*-}
& 0 & 0 & 0 & 0 
& 0 & 0 
& -{1 \over 3}  & 0  & 0 & -{3 \over 2} 
& {1 \over {3}}  & -{1 \over 4}  
& -{1 \over 3} & 0 & -{1 \over 4} 
\\
 {1 \over \sqrt{2}} \Xi\Xi^{*0}
& {1 \over 3} & 0 & 0 & {3 \over 2} 
& 0 & 0 
& 0 & 0  & 0 & 0 
& -{1 \over {3}}  & {1 \over 4}  
& -{1 \over 3} & 0 & -{5 \over {12}} 
\\
 {1 \over \sqrt{2}} \Xi\Xi^{*-}
& 0 & 0 & 0 & 0 
& 0 & 0 
& -{1 \over 3}  & 0  & 0 & -{3 \over 2}  
& {1 \over {3}}  & -{1 \over 4}  
& -{1 \over 3} & 0 & -{1 \over 4} 
\\[1pt]
%
%
%
%%
%}}}
\hline
\end{array}
\end{eqnarray*}
\end{table*}

\setlength{\arraycolsep}{3pt}
\renewcommand{\arraystretch}{1.7}

\begin{table*}[htbp]
\caption{Matrix Elements of Subleading $SU(3)$-Violating and Isospin-Violating Magnetic Moment Operators.  The subleading 
$SU(3)$-violating operators are the first six operators, and the isospin-violating operators are the second six operators. }
\smallskip
\label{tab:12ops}
\begin{eqnarray*}
\begin{array}{|c|c|c|c|c|c|c||c|c|c|c|c|c|c|c|c|}
\hline
& {1 \over \sqrt{3}} O_{(2)}^{i(8\overline Q)} & {1 \over \sqrt{3}} D_{(3)}^{i(8 \overline Q)} 
& {1 \over \sqrt{3}} O_{(3)}^{i(8 \overline Q)} 
& O_{(3)}^{i(88Q)}  & \tilde O_{(3)}^{i (88Q)}  & O_{(3)}^{i(88 \overline Q)}
& O_{(2)}^{i(3Q)} &  D_{(3)}^{i(3Q)}  &  O_{(3)}^{i(3Q)}
& O_{(3)}^{i(33Q)}  & \tilde O_{(3)}^{i (33Q)}  & O_{(3)}^{i(33 \overline Q)} \\[2pt]
\hline
& (1, {\bf 27}) & (1, {\bf 27}) & (1, {\bf 27}) & (1, {\bf 64}) & (1, {\bf 64}) & (1, {\bf 64}) & (1, {\bf 27}) & (1, {\bf 27}) & (1, {\bf 27}) 
& (1, {\bf 64}) 
& (1, {\bf 64}) & (1, {\bf 64})  \\
\hline
p
& {1 \over 3} & 0 & {1 \over 6} 
& {1 \over 2} & {3 \over 2}  & 0 
& {4 \over 3}  &  {1 \over 2} & {5 \over 4} 
& {5 \over 6}  &  {1 \over 2} & 0 
\\
 n
& -{2 \over 3} & -{1 \over 2} & -{1 \over 4} 
&  0  & -1 & -{1 \over 2} 
& {1 \over 3}  & 0  & {5 \over 6} 
& 0  & -{1 \over 3}  & -{5 \over 6} 
\\
 \Lambda
& 0 & 0 & -{1 \over 6} 
& 0   &  0 & 0 
&  0 & 0  & {1 \over 2} 
& 0  &  0 & 0 
\\
 \Lambda \Sigma^0
& 0 & 0 & 0 
& 0   &  0 & 0 
&  0 & 0  & 0 
& 0  &  0 & 0 
\\
 \Sigma^+
& {1 \over 3} & 0 & {1 \over 6} 
& 0   &  0 & 0 
& {5 \over 3}  & 1  & 1 
& {4 \over 3}  &  2 & {4 \over 3} 
\\
 \Sigma^0
& 0 & 0 & -{1 \over 6} 
& 0   &  0 & 0 
& 0 & 0  & {1 \over 2} 
& 0  &  0 & 0 
\\
 \Sigma^-
& -{1 \over 3} & 0 & -{1 \over 2} 
& 0   &  0 & 0 
& 1  &  1 & {1 \over 3} 
& -{4 \over 3}  &  -{2 \over 3} & -{4 \over 3} 
\\
 \Xi^0
& -{2 \over 3} & -{1 \over 2} & -{1 \over 4} 
&  0  &  -1 & {3 \over 2} 
& -{1 \over 3}  & 0  & {1 \over 6} 
& 0  & -{1 \over 3}  & -{1 \over 6} 
\\
 \Xi^-
& -{1 \over 3} & 0 & -{1 \over 2} 
&  -{3 \over 2}  & - {1 \over 2}  & 0 
& 0  & {1 \over 2}  & -{1 \over {12}} 
& {1 \over 6}  & -{1 \over 6}  & 0 
\\
 \Delta^{++}
& 1 & {3 \over 2} & {5 \over 4} 
&  {3}  &  3  & {3 \over 2} 
& 6  & 9  & {{15} \over 2} 
&  9 & 9  & {9 \over 2} 
\\
 \Delta^{+}
& 0 & 0 & 0 
& {3 \over 2}   &  {3 \over 2}  & 0 
& 1  & {3 \over 2}  & {5 \over 4} 
& {1 \over 2}  & {1 \over 2}  & 0 
\\
 \Delta^{0}
& -1 & -{3 \over 2} & -{5 \over 4} 
&  0  &  0  & -{3 \over 2} 
& 0  & 0  & 0 
& 0  & 0  & -{1 \over 2} 
\\
 \Delta^{-}
& -2 & -3 & -{5 \over 2} 
&  -{3 \over 2}  & -{3 \over 2}  & -3 
& 3  & {9 \over 2}  & {{15} \over 4} 
& -{9 \over 2}  & -{9 \over 2}  & -9 
\\
 \Sigma^{*+}
& 0 & 0 & 0 
&  0  &  0  & 0 
& 2  &  3 & {5 \over 2} 
& 2  & 2  & 2 
\\
 \Sigma^{*0}
& 0 & 0 & 0 
&  0  &  0  & 0 
& 0  & 0  & 0 
& 0  &  0 & 0 
\\
 \Sigma^{*-}
& 0 & 0 & 0 
&  0  &  0  & 0 
& 2  & 3  & {5 \over 2} 
& -2  & -2  & -2 
\\
 \Xi^{*0}
& -1 & -{3 \over 2} & -{5 \over 4} 
& 0   &  0  & {3 \over 2} 
& 0  & 0  & 0 
& 0  &  0 & {1 \over 2} 
\\
 \Xi^{*-}
& 0 & 0 & 0 
&  -{3 \over 2}  & -{3 \over 2}  & 0 
& 1  & {3 \over 2}  & {5 \over 4} 
& -{1 \over 2}  & -{1 \over 2}  & 0 
\\
 \Omega^-
& -2 & -3 & -{5 \over 2} 
&  -6  &  -6  & 6 
& 0 & 0 & 0 
& 0  & 0  & 0 
\\
 {1 \over \sqrt{2}} p\Delta^+
& {1 \over 3} & 0 & {1 \over {4}} 
& 0  &  1   & 0 
&  1 &  0 & {{13} \over {12}} 
& {2 \over 3}  & {1 \over 3}  & 0 
\\
 {1 \over \sqrt{2}} n\Delta^0
& {1 \over 3} & 0 & {1 \over {4}} 
& 0  &  1   & 0 
& -{1 \over 3}  & 0  & -{7 \over {12}} 
&  0 & {1 \over 3}  & {2 \over 3} 
\\
 {1 \over \sqrt{2}} \Lambda \Sigma^{*0}
& 0 & 0 & -{1 \over {4\sqrt{3}}} 
&  0  & 0  & 0 
&  0 &  0 & -{1 \over {4\sqrt{3}} } 
& 0  & 0  & 0 
\\
 {1 \over \sqrt{2}} \Sigma\Sigma^{*+}
& {1 \over 3} & 0 & {1 \over {4}} 
&  0  & 0  & 0 
&  1 & 0  & {{11} \over {12}} 
&  {2 \over 3} &   {4 \over 3}& {2 \over 3} 
\\
 {1 \over \sqrt{2}} \Sigma\Sigma^{*0}
& 0 & 0 & -{1 \over {12}}  
&  0  & 0  & 0 
& 0  & 0  & -{1 \over 4} 
& 0  & 0  & 0 
\\
 {1 \over \sqrt{2}} \Sigma\Sigma^{*-}
& -{1 \over 3} & 0 & -{5 \over {12}}  
&  0  & 0  & 0 
& {1 \over 3}  & 0  & {1 \over {4}} 
& -{2 \over 3}  &  0 & -{2 \over 3} 
\\
 {1 \over \sqrt{2}} \Xi\Xi^{*0}
& {1 \over 3} & 0 & {1 \over {4}} 
&  0  &  1 & -1 
& {1 \over 3}  & 0  & {1 \over {12}} 
& 0  &  {1 \over 3} & {1 \over 3} 
\\
 {1 \over \sqrt{2}} \Xi\Xi^{*-}
& {1 \over 3} & 0 & {5 \over {12}} 
& 1   &  0  & 0 
& {1 \over 3}  &  0 & {1 \over {4}} 
& -{1 \over 3}  &  0 & 0 
\\
\hline
\end{array}
\end{eqnarray*}
\end{table*}

\begin{table*}[htbp]
\caption{Coefficients of Magnetic Moment Operators Including $SU(3)$ Flavor Symmetry Violation at Leading Order}
\smallskip
\label{tab:coeffs15ops}
\begin{eqnarray*}
\begin{array}{|c|c|c|c||c|c|c|c|c|c|c|c|c|c|c|c|}
\hline
 {\rm Coefficient} & {\rm Operator}
& {\rm Magnetic \ Moment \ Combination} \\
\hline
 a_{(1)}^{1, {\bf 8}} & G^{iQ}
& {1 \over {32}} \left[ {{13}} N_1 - {{6\sqrt{3}}} \left(\Lambda \Sigma \right)_1 + 8 \Sigma_1 -{5 } \Xi_1 \right]
-{1 \over {60}} \left[ \Delta_1 + 2 \Sigma^*_1 + \Xi^*_1 \right] \\%
& &  
+{3 \over {20}} \left[  2 N_0 -3 \Lambda_0 +9 \Sigma_0 - 8 \Xi_0 \right]
-{1 \over {10}} \left[  2 \Delta_0 - \Xi_0^* -\Omega_0  \right]
\\
 b_{(2)}^{1, {\bf 8}} & J^i T^Q
& {1 \over {12}} \left[- N_1 + {{2\sqrt{3}}} \left(\Lambda \Sigma\right)_1 + {4} \Sigma_1 + {5 } \Xi_1 \right]
+{1 \over {5}} \left[ 7 N_0 +2 \Lambda_0 -6  \Sigma_0 -3 \Xi_0 \right]  
\\
 b_{(3)}^{1, {\bf 8}} & {\cal D}_{(3)}^{iQ}
& -{1 \over {96}} \left[ N_1 +{2\sqrt{3}} \left(\Lambda \Sigma\right)_1 +8 \Sigma_1 +7 \Xi_1 \right]
+ {1 \over {180}} \left[ \Delta_1 +2 \Sigma^*_1 +  \Xi^*_1 \right]
\\%
& & 
- {1 \over {20}} \left[ 6 N_0 + \Lambda_0 -3 \Sigma_0 -4 \Xi_0 \right]
+{1 \over {30}} \left[ 2 \Delta_0 - \Xi_0^* - \Omega_0 \right] 
\\
 c_{(3)}^{1, {\bf 8}} &  {\cal O}_{(3)}^{iQ}
& -{1 \over {144}} \left[ 13 N_1 -{6\sqrt{3}} \left(\Lambda \Sigma\right)_1 +8 \Sigma_1 -5 \Xi_1 \right]
+ {1 \over {270}} \left[ \Delta_1 + 2 \Sigma^*_1 +  \Xi^*_1 \right]
\\%
& & 
+ {2 \over {45}} {1 \over \sqrt{2}} \left[2 \left( N \Delta \right)_1 + {\sqrt{3}} \left( \Lambda \Sigma^{*} \right)_1
+   \left( \Sigma \Sigma^* \right)_1 +  \left( \Xi \Xi^* \right)_1 \right]
\\%
& & 
- {1 \over {30}} \left[ 2 N_0 -3 \Lambda_0 + 9 \Sigma_0 - 8 \Xi_0 \right] 
+{1 \over {45}} \left[ 2 \Delta_0 - \Xi_0^* - \Omega_0 \right]
+{2 \over {15}} {1 \over \sqrt{2}} \left[ 3 \left( \Sigma \Sigma^* \right)_0 
+ {2}  \left( \Xi \Xi^* \right)_0 \right]
\\
%
%
%%%%
%
 c_{(1)}^{1, {\bf 1}}  & J^i
&  {5 \over {16}} \left[ 2N_0 + \Lambda_0 + 3 \Sigma_0+ 2 \Xi_0  \right]  -{1 \over {60}} \left[ 4 \Delta_0 + 3 \Sigma^{*}_0  + 2 \Xi^{*}_0 + \Omega_0 \right]
\\
 c_{(3)}^{1, {\bf 1}}  & \{ J^2, J^i \}
&  -{1 \over {24}} \left[ 2 N_0 + \Lambda_0 + 3 \Sigma_0 + 2 \Xi_0  \right] 
+{1 \over {90}} \left[ 4 \Delta_0  + 3 \Sigma^{*}_0  + 2 \Xi^{*}_0 + \Omega_0 \right]
\\
%
%%%%%
%
 {1 \over \sqrt{3}} \overline a_{(1)}^{1, {\bf 8}} & G^{i\overline Q}
& {1 \over {32}} \left[{{13}} N_1 - {{6\sqrt{3}}} \left( \Lambda \Sigma \right)_1 + 8 \Sigma_1 -{5 } \Xi_1 \right]
-{1 \over {60}} \left[ \Delta_1 + 2 \Sigma^*_1 + \Xi^*_1 \right] \\%
& & 
-{3 \over {20}} \left[  2 N_0 -3 \Lambda_0 +9 \Sigma_0 - 8 \Xi_0 \right]
+{1 \over {10}} \left[  2 \Delta_0 - \Xi_0^* -\Omega_0 \right]
\\
 {1 \over \sqrt{3}} \overline b_{(2)}^{1, {\bf 8}} & J^i T^{\overline Q} 
& {1 \over {12}} \left[- N_1 + {{2\sqrt{3}}} \left( \Lambda \Sigma \right)_1 + {4} \Sigma_1 + {5 } \Xi_1 \right]
%\\%
%& & &
-{1 \over {5}} \left[ 7 N_0 +2 \Lambda_0 -6  \Sigma_0 -3 \Xi_0 \right]
\\
 {1 \over \sqrt{3}} \overline b_{(3)}^{1, {\bf 8}} & {\cal D}_{(3)}^{i \overline Q} 
& -{1 \over {96}} \left[ N_1 +{2\sqrt{3}} \left( \Lambda \Sigma \right)_1 +8 \Sigma_1 +7 \Xi_1 \right]
+ {1 \over {180}} \left[ \Delta_1 +2 \Sigma^*_1 +  \Xi^*_1 \right]
\\%
& & 
+ {1 \over {20}} \left[ 6 N_0 + \Lambda_0 -3 \Sigma_0 -4 \Xi_0 \right]
-{1 \over {30}} \left[ 2 \Delta_0 - \Xi_0^* - \Omega_0 \right] 
\\
 {1 \over \sqrt{3}} \overline c_{(3)}^{1, {\bf 8}}  & {\cal O}_{(3)}^{i \overline Q}  
& -{1 \over {144}} \left[ 13 N_1 -{6\sqrt{3}} \left( \Lambda \Sigma \right)_1 +8 \Sigma_1 -5 \Xi_1 \right]
+ {1 \over {270}} \left[ \Delta_1 + 2 \Sigma^*_1 +  \Xi^*_1 \right]
\\%
& & 
+ {2 \over {45}} {1 \over \sqrt{2}} \left[2 \left( N \Delta \right)_1 + {\sqrt{3}}  \Lambda \Sigma^{*0} 
+   \left( \Sigma \Sigma^* \right)_1 +  \left( \Xi \Xi^* \right)_1 \right]
\\%
& & 
+ {1 \over {30}} \left[ 2 N_0 -3 \Lambda + 9 \Sigma_0 - 8 \Xi_0 \right] 
-{1 \over {45}} \left[ 2 \Delta_0 - \Xi_0^* - \Omega  \right]
-{2 \over {15}} {1 \over \sqrt{2}} \left[ 3 \left( \Sigma \Sigma^* \right)_0 
+ {2}  \left( \Xi \Xi^* \right)_0 \right]
\\
%
%
%%%%%
%
 {1 \over \sqrt{3}} c_{(2)}^{1, {\bf 10 + \overline {10}}}  & O_{(2)}^{i[8Q]}
%\{ T^8, G^{iQ} \} - \{ G^{i8}, T^Q \}
& {1 \over 2} \left[N_1-\Sigma_1 + \Xi_1  \right]
\\
 {1 \over \sqrt{3}} c_{(3)}^{1, {\bf 10 + \overline {10}}}  & O_{(3)}^{i[8Q]}
%\{ G^{i8}, J^k G^{kQ} \} - \{ J^k G^{k8}, G^{iQ} \}
& {2 \over 3} \left[N_1-\Sigma_1 + \Xi_1  \right]
+ {2 \over 3} \left[- {1 \over \sqrt{2}}\left( N \Delta \right)_1 
+{1 \over \sqrt{2}} \left( \Sigma\Sigma^{*}\right)_1 
+{1 \over \sqrt{2}}\left( \Xi\Xi^{*} \right)_1 \right]
\\
%
%%%%%
%
 {1 \over \sqrt{3}} c_{(2)}^{1, {\bf 27}}  & O_{(2)}^{i(8Q)}
%\{ T^8, G^{iQ} \} + \{ G^{i8}, T^Q \}
&  {5 \over 6} \left[ N_1 + {2\sqrt{3}} \left( \Lambda \Sigma \right)_1 - \Xi_1 \right]
-{5 \over {126}} \left[ \Delta_1 -3 \Sigma^*_1 -4 \Xi^*_1 \right]
\\%
& & 
+ {1 \over 6} {1 \over \sqrt{2}} \left[ \left( N \Delta \right)_1  -{2\sqrt{3}}  \left( \Lambda \Sigma^* \right)_1
+ 3 \left( \Sigma \Sigma^* \right)_1 -2 \left( \Xi \Xi^* \right)_1\right] 
\\%
& & 
 +{5 \over 6} \left[ 2 N_0 -3 \Lambda_0 - \Sigma_0 + 2 \Xi_0 \right]
-{5 \over {63}} \left[ 4 \Delta_0 -5  \Sigma^*_0 -2 \Xi^*_0 +3 \Omega_0 \right]
\\%
& & 
+ {4 \over 3}{1 \over \sqrt{2}} \left[ \left( \Sigma \Sigma^* \right)_0 - \left( \Xi \Xi^* \right)_0 \right]
\\
 {1 \over \sqrt{3}} b_{(3)}^{1, {\bf 27}}   &  D_{(3)}^{i(8Q)}
%J^i \{ T^8, T^Q \}
& \quad {1 \over {21}} \left[  \Delta_1 -3 \Sigma^*_1 -4 \Xi^*_1 \right]
-{1 \over 4}{1 \over \sqrt{2}} \left[ \left( N \Delta \right)_1 - {2 \sqrt{3}}  \left( \Lambda \Sigma^* \right)_1
+3\left( \Sigma \Sigma^* \right)_1 - 2 \left( \Xi \Xi^* \right)_1\right] \quad
\\%
&  & 
+ {2 \over {21}} \left[ 4 \Delta_0 -5 \Sigma^*_0 -2 \Xi^*_0 + 3 \Omega_0 \right]
-2{1 \over \sqrt{2}} \left[ \left( \Sigma \Sigma^* \right)_0 -  \left( \Xi \Xi^* \right)_0 \right]
\\
 {1 \over \sqrt{3}} c_{(3)}^{1, {\bf 27}}   &  O_{(3)}^{i(8Q)}
%\{ G^{i8}, J^k G^{kQ} \} + \{ J^k G^{k8}, G^{iQ} \}
& -{2 \over 3} \left[ N_1 +{2 \sqrt{3}} \left( \Lambda \Sigma \right)_1- \Xi_1 \right]
+{2 \over {63}} \left[ \Delta_1 -3 \Sigma^*_1 -4 \Xi^*_1 \right]
\\%
& &
 + {1 \over 6} {1 \over \sqrt{2}} \left[ \left( N \Delta \right)_1  -{2\sqrt{3}}  \left( \Lambda \Sigma^* \right)_1
+ 3 \left( \Sigma \Sigma^* \right)_1 -2 \left( \Xi \Xi^* \right)_1\right] 
\\%
& &
 -{2 \over 3} \left[ 2 N_0 -3 \Lambda_0 - \Sigma_0 + 2 \Xi_0 \right]
+{4 \over {63}} \left[ 4 \Delta_0 -5  \Sigma^*_0 -2 \Xi^*_0 +3 \Omega_0 \right]
\\%
& & 
+ {4 \over 3}{1 \over \sqrt{2}} \left[ \left( \Sigma \Sigma^* \right)_0 - \left( \Xi \Xi^* \right)_0 \right]
\\
%
%
%
%%%
\hline
\end{array}
\end{eqnarray*}
\end{table*}

\begin{table*}[htbp]
\caption{Coefficients of Magnetic Moment Operators Breaking $SU(3)$ Flavor Symmetry at Subleading Order}
\smallskip
\label{tab:coeffssubleading}
\begin{eqnarray*}
\begin{array}{|c|c|c|c||c|c|c|c|c|c|c|c|c|c|c|c|}
\hline
%%%
{\rm Coefficient} & {\rm Operator}
& {\rm Magnetic \ Moment \ Combination} \\
\hline
 {1 \over {3}} \overline c_{(2)}^{1, {\bf 27}}   &   O_{(2)}^{i(8 \overline Q)}
%\left( \{ T^8, G^{i\overline Q} \} + \{ G^{i8}, T^{\overline Q} \} \right) 
&  {5 \over 6} \left[ N_1 + {2\sqrt{3}} \left( \Lambda \Sigma \right)_1 - \Xi_1 \right]
-{5 \over {126}} \left[ \Delta_1 -3 \Sigma^*_1 -4 \Xi^*_1 \right]
\\%
 & &
+ {1 \over 6} {1 \over \sqrt{2}} \left[ \left( N \Delta \right)_1  -{2\sqrt{3}}  \left( \Lambda \Sigma^* \right)_1
+ 3 \left( \Sigma \Sigma^* \right)_1 -2 \left( \Xi \Xi^* \right)_1\right] 
\\%
 & &
 -{5 \over 6} \left[ 2 N_0 -3 \Lambda_0 - \Sigma_0 + 2 \Xi_0 \right]
+{5 \over {63}} \left[ 4 \Delta_0 -5  \Sigma^*_0 -2 \Xi^*_0 +3 \Omega_0 \right]
-{4 \over 3} {1 \over \sqrt{2}} \left[ \left( \Sigma \Sigma^* \right)_0 - \left( \Xi \Xi^* \right)_0 \right]
\\
 {1 \over {3}} \overline b_{(3)}^{1, {\bf 27}}   &  D_{(3)}^{i(8 \overline Q)}
%J^i \{ T^8, T^{\overline Q} \} 
& \quad {1 \over {21}} \left[  \Delta_1 -3 \Sigma^*_1 -4 \Xi^*_1 \right]
-{1 \over 4}{1 \over \sqrt{2}} \left[ \left( N \Delta \right)_1 - {2 \sqrt{3}}  \left( \Lambda \Sigma^* \right)_1
+3\left( \Sigma \Sigma^* \right)_1 - 2 \left( \Xi \Xi^* \right)_1\right] \quad
\\%
  &   &
- {2 \over {21}} \left[ 4 \Delta_0 -5 \Sigma^*_0 -2 \Xi^*_0 + 3 \Omega_0 \right]
+2{1 \over \sqrt{2}} \left[ \left( \Sigma \Sigma^* \right)_0 -  \left( \Xi \Xi^* \right)_0 \right]
\\
 {1 \over {3}}  \overline c_{(3)}^{1, {\bf 27}}   &  O_{(3)}^{i(8 \overline Q)}
%\left( \{ G^{i8}, J^k G^{k\overline Q} \} + \{ J^k G^{k8}, G^{i\overline Q} \} \right) 
& -{2 \over 3} \left[ N_1 +{2 \sqrt{3}} \left( \Lambda \Sigma \right)_1- \Xi_1 \right]
+{2 \over {63}} \left[ \Delta_1 -3 \Sigma^*_1 -4 \Xi^*_1 \right]
\\%
 & &
 + {1 \over 6} {1 \over \sqrt{2}} \left[ \left( N \Delta \right)_1  -{2\sqrt{3}}  \left( \Lambda \Sigma^* \right)_1
+ 3 \left( \Sigma \Sigma^* \right)_1 -2 \left( \Xi \Xi^* \right)_1\right] 
\\%
 & &
 +{2 \over 3} \left[ 2 N_0 -3 \Lambda_0 - \Sigma_0 + 2 \Xi_0 \right]
-{4 \over {63}} \left[ 4 \Delta_0 -5  \Sigma^*_0 -2 \Xi^*_0 +3 \Omega_0 \right]
\\%
 & &
- {4 \over 3}{1 \over \sqrt{2}} \left[ \left( \Sigma \Sigma^* \right)_0 - \left( \Xi \Xi^* \right)_0 \right]
\\
%
%
%
%%%%%
%
 c_{(3)}^{1, {\bf 64}}   & O_{(3)}^{i(88Q)} 
%\{ G^{i8},\{T^8, T^Q \} \}
& {1 \over {45}} \left[ \Delta_1 - 10 \Sigma^*_1 + 10 \Xi^*_1 \right]
+ {1 \over 6}{1 \over \sqrt{2}} \left[
- \left( N\Delta \right)_1
+{2\sqrt{3}} \left( \Lambda \Sigma^{*} \right)_1
 + \left( \Sigma \Sigma^{*}  \right)_1
 -2 \left( \Xi \Xi^{*}  \right)_1 \right]
 \\%
  & &
+ {2 \over 9} \left[ \Delta_0 -3 \Sigma^*_0 + 3 \Xi^*_0 - \Omega_0 \right]
 \\
 d_{(3)}^{1, {\bf 64}}    & \tilde O_{(3)}^{i(88Q)} 
%\{ T^8, \{ T^8, G^{iQ} \} \}
&{1 \over {90}} \left[ \Delta_1 - 10 \Sigma^*_1 + 10 \Xi^*_1 \right]
-{1 \over 6}{1 \over \sqrt{2}} \left[ -\left( N \Delta \right)_1 +2 \sqrt{3} \left( \Lambda \Sigma^* \right)_1 + \left( \Sigma \Sigma^* \right)_1 - 2 \left( \Xi \Xi^* \right)_1 \right]
\\
 {1 \over \sqrt{3}} \overline c_{(3)}^{1, {\bf 64}}   & O_{(3)}^{i(88 \overline Q)} 
%\{ G^{i8}, \{ T^8, T^{\overline Q} \} \}  
& {1 \over {45}} \left[ \Delta_1 - 10 \Sigma^*_1 + 10 \Xi^*_1 \right]
+ {1 \over 6}{1 \over \sqrt{2}} \left[
- \left( N\Delta \right)_1
+{2\sqrt{3}} \left( \Lambda \Sigma^{*} \right)_1
 + \left( \Sigma \Sigma^{*}  \right)_1
 -2 \left( \Xi \Xi^{*}  \right)_1 \right]
\\%
 & &
 -{2 \over 9} \left[ \Delta_0 -3 \Sigma^*_0 + 3 \Xi^*_0 - \Omega_0 \right]
\\
%
%
%%%
\hline
\end{array}
\end{eqnarray*}
\end{table*}

\begin{table*}[htbp]
\caption{Coefficients of Magnetic Moment Operators Violating Isospin Symmetry}
\smallskip
\label{tab:coeffsiso}
\begin{eqnarray*}
\begin{array}{|c|c|c|c||c|c|c|c|c|c|c|c|c|c|c|c|}
\hline
 {\rm Coefficient} & {\rm Operator}
& {\rm Magnetic \ Moment \ Combination} \\
\hline
    g_{(2)}^{1, {\bf 27}}  & O_{(2)}^{i(3Q)}
%\left(\{ T^3, G^{iQ} \} + \{ G^{i3}, T^Q \} \right)
&  {{5} \over {12}} \Sigma_2 -{5 \over {252}} \left[3 \Delta_2 +  \Sigma^*_2 \right]
+{1 \over {12}} {1 \over \sqrt{3}} \left[ 3 \left( N \Delta \right)_2 + \left(\Sigma \Sigma^{*}\right)_2 \right]
\\
 f_{(3)}^{1, {\bf 27}}  & D_{(3)}^{i(3Q)}
%J^i \{ T^3, T^Q \} 
& {1 \over {42}} \left[ 3 \Delta_2 +  \Sigma^*_2 \right]
-{1 \over 8} {1 \over \sqrt{2}} \left[ 3 \left( N \Delta \right)_2  + \left( \Sigma \Sigma^*\right)_2 \right]
\\
  g_{(3)}^{1, {\bf 27}}  & O_{(3)}^{i(3Q)}
%\left( \{ G^{i3} , J^k G^{kQ} \} + \{ J^k G^{k3} , G^{iQ} \} \right) 
&   -{1 \over 3} \Sigma_2 +{1 \over {63}} \left[ 3 \Delta_2+ \Sigma^*_2 \right] 
+ {1 \over {12}} {1 \over \sqrt{2}} \left[3 \left( N \Delta \right)_2 + \left( \Sigma \Sigma^{*} \right)_2 \right]
\\
%
%%%
%
  g_{(3)}^{1, {\bf 64}} & O_{(3)}^{i(33Q)}
%\{ G^{i3}, \{ T^3, T^Q \} \} 
& {1 \over {12}} \Delta_3 
+ {1 \over {6}}\left[ \Delta_2 -2 \Sigma^{*}_2 \right] 
+{1 \over 2} {1 \over \sqrt{2}} \left[ \left( N \Delta \right)_2 - \left( \Sigma \Sigma^{*} \right)_2 \right]
\\
 f_{(3)}^{1, {\bf 64}} & \tilde O_{(3)}^{i(33Q)}
%\{ T^3, \{ T^3, G^{iQ} \} \} 
&  {1 \over {12}}\left[ \Delta_2 -2 \Sigma^{*}_2 \right] 
-{1 \over 2} {1 \over \sqrt{2}} \left[ \left( N \Delta \right)_2 - \left( \Sigma \Sigma^{*} \right)_2 \right]
\\
 \overline g_{(3)}^{1, {\bf 64}} & O_{(3)}^{i(33 \overline Q)}
%\{ G^{i3}, \{ T^3, T^{\overline Q} \} \} 
&  
{1 \over {12}} \Delta_3
- {1 \over {6}}\left[ \Delta_2 -2 \Sigma^{*}_2 \right] 
-{1 \over 2} {1 \over \sqrt{2}} \left[ \left( N \Delta \right)_2 - \left( \Sigma \Sigma^{*} \right)_2 \right]
\\
\hline
\end{array}
\end{eqnarray*}
\end{table*}

\begin{table*}[htbp]
\caption{Coefficients of $(1, {\bf 27})$ Magnetic Moment Operators with Isospin $I=0$, $1$, and $2$.}
\smallskip
\label{tab:coeffs27}
\begin{eqnarray*}
\begin{array}{|c|c|c|}
\hline
 {\rm Coefficient} & {\rm Operator}
& {\rm Magnetic \ Moment \ Combination} \\
\hline
 \left( c_{(2)}^{1, {\bf 27}} \right)_{I=0}  &  \{ G^{i8}, T^8 \} 
&  {5 \over 6} \left[ 2 N_0 -3 \Lambda_0 - \Sigma_0 + 2 \Xi_0 \right]
-{5 \over {63}} \left[ 4 \Delta_0 -5  \Sigma^*_0 -2 \Xi^*_0 +3 \Omega_0 \right]
\\%
 & &
+ {4 \over 3}{1 \over \sqrt{2}} \left[ \left( \Sigma \Sigma^* \right)_0 - \left( \Xi \Xi^* \right)_0 \right]
\\ %%
 \left( c_{(3)}^{1, {\bf 27}} \right)_{I=0}  &  \{ G^{i8}, J^k G^{k8} \}
&  -{2 \over 3} \left[  2 N_0 - {3} \Lambda_0 -  \Sigma_0 +2 \Xi_0 \right]
+ {4 \over {63}} \left[ 4 \Delta_0 -5 \Sigma^*_0 -2 \Xi^*_0 + 3 \Omega_0 \right]
\\%
 & &
+ {4 \over 3}{1 \over \sqrt{2}} \left[ \left( \Sigma \Sigma^* \right)_0 - \left( \Xi \Xi^* \right)_0 \right]
\\%%
 \left( b_{(3)}^{1, {\bf 27}} \right)_{I=0}  & J^i \{ T^8, T^8 \} 
&  {2 \over {21}} \left[ 4 \Delta_0 -5 \Sigma^*_0 -2 \Xi^*_0 + 3 \Omega_0 \right]
-2{1 \over \sqrt{2}} \left[ \left( \Sigma \Sigma^* \right)_0 -  \left( \Xi \Xi^* \right)_0 \right]
\\
%
%
%%%%%
%
 \left( c_{(2)}^{1, {\bf 27}} \right)_{I=1}  &  \{ G^{i3}, T^8 \} + \{ G^{i8} , T^3 \}
&  {5 \over 6} \left[ N_1 + {2\sqrt{3}} \left( \Lambda \Sigma \right)_1 - \Xi_1 \right]
-{5 \over {126}} \left[ \Delta_1 -3 \Sigma^*_1 -4 \Xi^*_1 \right]
\\%
 & &
+ {1 \over 6} {1 \over \sqrt{2}} \left[ \left( N \Delta \right)_1  -{2\sqrt{3}}  \left( \Lambda \Sigma^* \right)_1
+ 3 \left( \Sigma \Sigma^* \right)_1 -2 \left( \Xi \Xi^* \right)_1\right] 
\\ %%
 \left( c_{(3)}^{1, {\bf 27}} \right)_{I=1}  &  \{ G^{i3}, J^k G^{k8} \} + \{ G^{i8}, J^k G^{k3} \}
& -{2 \over 3} \left[ N_1 +{2 \sqrt{3}} \left( \Lambda \Sigma \right)_1-  \Xi_1 \right]
+{2 \over {63}} \left[ \Delta_1 -3 \Sigma^*_1 -4 \Xi^*_1 \right]
\\%
 & &
+{1 \over 6}{1 \over \sqrt{2}} \left[ \left( N \Delta \right)_1-{2 \sqrt{3}} \left( \Lambda \Sigma^* \right)_1
+ 3 \left( \Sigma \Sigma^*\right)_1 -2 \left( \Xi \Xi^* \right)_1 \right]
\\%%
 \left( b_{(3)}^{1, {\bf 27}} \right)_{I=1}  &  J^i \{ T^8, T^3 \} 
& \quad {1 \over {21}} \left[  \Delta_1 -3 \Sigma^*_1 -4 \Xi^*_1 \right]
-{1 \over 4}{1 \over \sqrt{2}} \left[ \left( N \Delta \right)_1 - {2 \sqrt{3}}  \left( \Lambda \Sigma^* \right)_1
+3\left( \Sigma \Sigma^* \right)_1 - 2 \left( \Xi \Xi^* \right)_1\right] \quad
\\
%
%
%%%%%
%
 \left( c_{(2)}^{1, {\bf 27}} \right)_{I=2}   &  \{ G^{i3}, T^3 \}
& {5 \over {12}} \Sigma_2 -{5 \over {252}} \left[3 \Delta_2  + \Sigma^*_2 \right]
+ {1 \over {12}} {1 \over \sqrt{2}} \left[3 \left( N \Delta \right)_2 +  \left( \Sigma \Sigma^* \right)_2 \right]
\\%%
 \left( c_{(3)}^{1, {\bf 27}} \right)_{I=2}   &  \{ G^{i3}, J^k G^{k3}  \}
& -{1 \over 3} \Sigma_2 + {1 \over {63}} \left[  3 \Delta_2 + \Sigma^*_2 \right]
+ {1 \over {12}} {1 \over \sqrt{2}} \left[ 3 \left( N \Delta \right)_2 + \left( \Sigma \Sigma^* \right)_2 \right]
\\%%
 \left( b_{(3)}^{1, {\bf 27}} \right)_{I=2}   &  J^i \{ T^3, T^3 \}
& {1 \over {42}} \left[ 3 \Delta_2 +  \Sigma^*_2 \right]
-{1 \over 8}{1 \over \sqrt{2}}\left[ 3 \left( N \Delta \right)_2 + \left( \Sigma \Sigma^* \right)_2 \right]
\\[5pt]
%
%
%%%
\hline
\end{array}
\end{eqnarray*}
\end{table*}

\begin{table*}[htbp]
\caption{Coefficients of $(1, {\bf 64})$ Magnetic Moment Operators with Isospin $I=0$, $1$, $2$, and $3$.}
\smallskip
\label{tab:coeffs64}
\begin{eqnarray*}
\begin{array}{|c|c|c|}
\hline
 {\rm Coefficient} & {\rm Operator}
& {\rm Magnetic \ Moment \ Combination} \\
\hline
 \left( c_{(3)}^{1, {\bf 64}} \right)_{I=0}   &  \{ G^{i8},\{T^8, T^8 \} \}
& {2 \over {9}} \left[ \Delta_0 - 3 \Sigma^*_0 + 3 \Xi^*_0 - \Omega_0 \right] 
\\
%
%
%%%%%
%
 \left( c_{(3)}^{1, {\bf 64}} \right)_{I=1}  &  \{ G^{i3},\{T^8, T^8 \} \}
&  \quad {1 \over {90}} \left[ \Delta_1 -{10} \Sigma^*_1 + {10} \Xi^*_1 \right]
- {1 \over 6} {1 \over \sqrt{2}} \left[
-\left( N \Delta \right)_1 + {2 \sqrt{3}} \left( \Lambda \Sigma^* \right)_1+ 
 \left( \Sigma \Sigma^* \right)_1 -2 \left( \Xi \Xi^* \right)_1 \right] \quad
\\ %%
 \left( c_{(3)}^{1, {\bf 64}} \right)_{I=1}  &  \{ G^{i8},\{T^8, T^3 \} \}
&  \quad {1 \over {45}} \left[ \Delta_1 -10 \Sigma^*_1 + 10 \Xi^*_1 \right]
+ {1 \over 6} {1 \over \sqrt{2}} \left[
- \left( N \Delta \right)_1 + {2 \sqrt{3}} \left( \Lambda \Sigma^* \right)_1+ 
 \left( \Sigma \Sigma^* \right)_1 -2 \left( \Xi \Xi^* \right)_1 \right] \quad
\\
%
%
%%%%%
%
 \left( c_{(3)}^{1, {\bf 64}} \right)_{I=2}  &  \{ G^{i3},\{T^8, T^3 \} \}
&  {1 \over 6} \left[ \Delta_2 -2 \Sigma^*_2 \right]
+ {1 \over 2} {1 \over \sqrt{2}}\left[ \left( N \Delta \right)_2 -\left( \Sigma \Sigma^* \right)_2 \right]
\\ %%
 \left( c_{(3)}^{1, {\bf 64}} \right)_{I=2}  &  \{ G^{i8},\{T^3, T^3 \} \}
&  {1 \over {12}} \left[ \Delta_2 -2 \Sigma^*_2 \right]
-{1 \over 2} {1 \over \sqrt{2}} \left[ \left( N \Delta \right)_2 -\left( \Sigma \Sigma^* \right)_2 \right]
\\
%
%%%%%
%
 \left( c_{(3)}^{1, {\bf 64}} \right)_{I=3}   &  \{ G^{i3}, \{ T^3, T^3 \} \}
& {1 \over {12}} \Delta_3
\\[5pt]
%
%
%
%%%
\hline
\end{array}
\end{eqnarray*}
\end{table*}

\begin{table*}[htbp]
\caption{Coefficients of Magnetic Moment Operators to all orders in $SU(3)$ Flavor Symmetry Breaking.  The first four operators contribute in the $SU(3)$ flavor symmetry limit.  There are 21 operators.}
\smallskip
\label{tab:coeffssu3}
\begin{eqnarray*}
\begin{array}{|c|c|c|c|}
\hline
 {\rm Coefficient} & {\rm Operator}
& N_c & SU(3) \ {\rm Breaking} \\
\hline
 a_{(1)}^{1, {\bf 8}} & G^{iQ}
& N_c
& 1, \ m_q^{1/2}, \ m_q \ln m_q \\
 b_{(2)}^{1, {\bf 8}} & {1 \over N_c} J^i T^Q
& {1 \over N_c}
& 1, \ m_q^{1/2}, \ m_q \ln m_q   \\
 b_{(3)}^{1, {\bf 8}} & {1 \over N_c^2} {\cal D}_{(3)}^{iQ}
& {1 \over N_c}
& 1, \ m_q^{1/2}, \ m_q \ln m_q    \\
 c_{(3)}^{1, {\bf 8}} &  {1 \over N_c^2} {\cal O}_{(3)}^{iQ}
& {1 \over N_c}
& 1, \ m_q^{1/2}, \ m_q \ln m_q   \\
%
%%%
%
 c_{(1)}^{1, {\bf 1}}  & J^i
& 1 & m_q \ln m_q \\
 c_{(3)}^{1, {\bf 1}}  & {1 \over N_c^2} \{ J^2, J^i \}
& {1 \over N_c^2} & m_q \ln m_q \\
%
%%%
%
 \overline a_{(1)}^{1, {\bf 8}} & G^{i\overline Q}
& N_c & m_q^{1/2} \\
 \overline b_{(2)}^{1, {\bf 8}} & {1 \over N_c} J^i T^{\overline Q} 
& {1 \over N_c} & m_q^{1/2} \\
 \overline b_{(3)}^{1, {\bf 8}} & {1 \over N_c^2} {\cal D}_{(3)}^{i \overline Q} 
& {1 \over N_c} & m_q^{1/2} \\
 \overline c_{(3)}^{1, {\bf 8}}  & {1 \over N_c^2} {\cal O}_{(3)}^{i \overline Q}  
& {1 \over N_c} & m_q^{1/2} \\
%
%%%
%
 c_{(2)}^{1, {\bf 10 + \overline {10}}}  & {1 \over N_c} \left( \{ T^8, G^{iQ} \} - \{ G^{i8}, T^Q \} \right)
& 1 & m_q^{1/2} \\
 c_{(3)}^{1, {\bf 10 + \overline {10}}}  & {1 \over N_c^2} \left( \{ G^{i8}, J^k G^{kQ} \} - \{ J^k G^{k8}, G^{iQ} \} \right)
& {1 \over N_c} & m_q^{1/2} \\
%
%%%
%
 c_{(2)}^{1, {\bf 27}}  & {1 \over N_c} \left( \{ T^8, G^{iQ} \} + \{ G^{i8}, T^Q \} \right)
& 1 & m_q \ln m_q \\
 b_{(3)}^{1, {\bf 27}}   &  {1 \over N_c^2} J^i \{ T^8, T^Q \}
& {1 \over N_c^2} & m_q \ln m_q \\
 c_{(3)}^{1, {\bf 27}}   &  {1 \over N_c^2} \left( \{ G^{i8}, J^k G^{kQ} \} + \{ J^k G^{k8}, G^{iQ} \} \right)
& {1 \over N_c} & m_q \ln m_q  \\
%
%%%
%
 \overline c_{(2)}^{1, {\bf 27}}   &  {1 \over N_c} \left( \{ T^8, G^{i\overline Q} \} + \{ G^{i8}, T^{\overline Q} \} \right) 
& 1 &  2\ {\rm loop}\ \chi{\rm PT}
\\
 \overline b_{(3)}^{1, {\bf 27}}   &  {1 \over N_c^2} J^i \{ T^8, T^{\overline Q} \} 
& {1 \over N_c^2} & 2\ {\rm loop}\ \chi{\rm PT} \\
 \overline c_{(3)}^{1, {\bf 27}}   &  {1 \over N_c^2} \left( \{ G^{i8}, J^k G^{k\overline Q} \} + \{ J^k G^{k8}, G^{i\overline Q} \} \right) 
& {1 \over N_c} & 2\ {\rm loop}\ \chi{\rm PT} \\
%
%%%
%
 c_{(3)}^{1, {\bf 64}}   &  {1 \over N_c^2} \{ G^{i8},\{T^8, T^Q \} \}
& {1 \over N_c^2} & 2\ {\rm loop}\ \chi{\rm PT}  \\
 d_{(3)}^{1, {\bf 64}}    &  {1 \over N_c^2} \{ T^8, \{ T^8, G^{iQ} \} \}
& {1 \over N_c} & 2\ {\rm loop}\ \chi{\rm PT}   \\
 \overline c_{(3)}^{1, {\bf 64}}   &  {1 \over N_c^2} \{ G^{i8}, \{ T^8, T^{\overline Q} \} \}  
& {1 \over N_c^2}  & 3\ {\rm loop}\ \chi{\rm PT} \\
\hline
\end{array}
\end{eqnarray*}
\end{table*}

\begin{table*}[htbp]
\caption{Coefficients of Magnetic Moment Operators Violating Isospin Symmetry.  There are six operators.}
\smallskip
\label{tab:coeffsisobreaking}
\begin{eqnarray*}
\begin{array}{|c|c|c|c|}
\hline
 {\rm Coefficient} & {\rm Operator}
& N_c & \text{Isospin Breaking} \\
\hline
 g_{(2)}^{1, {\bf 27}}  & {1 \over N_c^2} \left( \{ T^3, G^{iQ} \} + \{ G^{i3}, T^Q \} \right)
& {1 \over N_c}  & m_q \ln m_q
\\
 f_{(3)}^{1, {\bf 27}}   &  {1 \over N_c^2} J^i \{ T^3, T^Q \}
& {1 \over N_c^2} & m_q \ln m_q
\\
 g_{(3)}^{1, {\bf 27}}  &  {1 \over N_c^2} \left( \{ G^{i3}, J^k G^{kQ} \} + \{ J^k G^{k3}, G^{iQ} \} \right)
& {1 } & m_q \ln m_q 
\\
%
%%%
%
  g_{(3)}^{1, {\bf 64}}  &  {1 \over N_c^2} \{ G^{i3},\{T^3, T^Q \} \}
& {1 \over N_c} & 2\ {\rm loop}\ \chi{\rm PT} 
\\
  f_{(3)}^{1, {\bf 64}}    &  {1 \over N_c^2} \{ T^3, \{ T^3, G^{iQ} \} \}
& {1 \over N_c} & 2\ {\rm loop}\ \chi{\rm PT}  
\\
 \overline g_{(3)}^{1, {\bf 64}}  &  {1 \over N_c^2} \{ G^{i3}, \{ T^3, T^{\overline Q} \} \}  
& {1 \over N_c}  & 3\ {\rm loop}\ \chi{\rm PT}
\\
\hline
\end{array}
\end{eqnarray*}
\end{table*}

\setlength{\arraycolsep}{1pt}
\renewcommand{\arraystretch}{1.4}

\begin{table*}[htbp]
\caption{Fits to Experimental Data.}
\smallskip
\label{tab:magmom_ops}
\begin{eqnarray*}
\begin{array}{|c|c|c|c|c|c|c|c|c|c|c|c|}
\hline
& {\rm Experiment } & {\rm Fit\ A}  & {\rm Fit\ A^\prime} & {\rm Fit\ B} & {\rm Fit\ B^\prime} & {\rm Fit\ C}  & {\rm Fit\ C^\prime} & {\rm Fit\ D} & {\rm Fit \ D^\prime}  
\\
\hline
 p
& 2.79 \pm 0.00 & 2.56  & 2.56 & 2.72 & 2.53 & 2.84 & 2.72 
& 2.91  & 2.83   
\\
 n
& -1.91 \pm 0.00 & -1.63  & -1.63 & -1.82 & -1.69 & -2.15 & -2.06 
& -2.22  & -2.17    
\\
 \Lambda
& -0.613 \pm 0.004 & -0.82  & -0.82 & -0.91 & -0.84 & -0.69 & -0.66 
& -0.69  & -0.66    
\\
 \Lambda \Sigma^0
& -1.61 \pm 0.08 & -1.41  & -1.41 & -1.57 & -1.46 & -1.73 & -1.66 
&  -1.73 & -1.64     
\\
 \Sigma^+
& 2.458 \pm 0.010 & 2.56  & 2.56 & 2.72 & 2.53 & 2.69 & 2.57 
&  2.62 & 2.42     
\\
 \Sigma^0
&  & 0.82  & 0.82 & 0.91 & 0.84 & 0.69 & 0.66 
& 0.69  & 0.66     
\\
 \Sigma^-
& -1.160 \pm 0.025 & -0.93  & -0.93 & -0.91 & -0.84 & -1.31 & -1.25 
& -1.23  &  -1.10     
\\
 \Xi^0
& -1.250 \pm 0.014 & -1.63  & -1.63 & -1.82 & -1.69 & -1.53 & -1.47 
& -1.47  &  -1.33     
\\
 \Xi^-
& -0.6507 \pm 0.0025 & -0.93  & -0.93 & -0.91 & -0.84 & -0.54 & -0.51 
& -0.61  & -0.65     
\\
 \Delta^{++}
& 6.14 \pm 0.51^\prime  & 5.23  & 4.04 & 5.45 & 5.07 & 5.53  & 5.29 
& 5.53  &  5.25    
\\
 \Delta^{+}
&  &  2.61 & 2.02  & 2.72 & 2.53 & 2.53 & 2.42 
& 2.54 &  2.41      
\\
 \Delta^{0}
&  & 0 & 0 & 0 & 0  & -0.46 & -0.44 
& -0.45 &  -0.43      
\\
 \Delta^{-}
&  & -2.61  & -2.02 & -2.72 & -2.53 & -3.46 & -3.31 
& -3.44 &  -3.27      
\\
 \Sigma^{*+}
&  & 2.61  & 2.02 & 2.72 & 2.53 & 3.00 & 2.87 
& 2.99 &  2.84      
\\
 \Sigma^{*0}
&  & 0  & 0 & 0 & 0 & 0 & 0 
& 0 & 0       
\\
 \Sigma^{*-}
&  &  -2.61 & -2.02 & -2.72 & -2.53 & -3.00 & -2.87 
&  -2.99 & -2.84      
\\
 \Xi^{*0}
&  & 0  & 0 & 0 & 0 & 0.46 & 0.44 
& 0.45  & 0.43     
\\
 \Xi^{*-}
&  & -2.61  & -2.02 & -2.72 & -2.53 & -2.53 & -2.42 
& -2.54  &  -2.41      
\\
 \Omega^-
&  -2.02 \pm 0.05 &  -2.61 & -2.02 & -2.72 & -2.53 & -2.07 & -1.98 
&  -2.08 &  -1.98       
\\
 {1 \over \sqrt{2}} p\Delta^+
& 2.48 \pm 0.06 & 2.36  & 2.48 & 1.82 & 1.69 & 2.00 & 1.91 
&  2.07 &  2.03     
\\
 {1 \over \sqrt{2}} n\Delta^0
&  & 2.36  & 2.48 & 1.82 & 1.69 & 2.00 & 1.91 
& 2.07  &  2.03      
\\
 {1 \over \sqrt{2}} \Lambda \Sigma^{*0}
& 1.94 \pm 0.18^\prime  & 2.04  & 2.15 & 1.57 & 1.46 & 1.73 & 1.66 
& 1.73 &  1.64      
\\
 {1 \over \sqrt{2}} \Sigma\Sigma^{*+}
&  2.28 \pm 0.25^\prime &  2.36 & 2.48 & 1.82 & 1.69 & 1.69 & 1.62 
& 1.62 &  1.47       
\\
 {1 \over \sqrt{2}} \Sigma\Sigma^{*0}
&  &  1.18 & 1.24 & 0.91 & 0.84 & 0.69 & 0.66 
& 0.69 &  0.66       
\\
 {1 \over \sqrt{2}} \Sigma\Sigma^{*-}
&  & 0 & 0 & 0 & 0 & -0.31 & -0.30 
& -0.23 &  -0.15      
\\
 {1 \over \sqrt{2}} \Xi\Xi^{*0}
&  & 2.36  & 2.48 & 1.82 & 1.69 & 1.69 & 1.62 
& 1.62 &  1.47      
\\
 {1 \over \sqrt{2}} \Xi\Xi^{*-}
&  &  0  & 0 & 0 & 0 & -0.31 & -0.30 
& -0.23 &  -0.15       
\\
%
%%%
%
 \sigma^{\rm theory}
& & 0.35  & 0.28 & 0.41 & 0.39 & 0.27 & 0.23 & 0.275 & 0.20     \\
 {\rm dof}
& & 9  & 6 & 12 & 9 & 11 & 8 & 10  & 7     \\
 a_{(1)}^{1,{\bf 8}}
&  & 4.98 \pm 0.49   & 5.28 \pm 0.40 & 5.45 \pm 0.31 & 5.07 \pm 0.35 & 5.07 \pm 0.23 & 4.85 \pm 0.21  
&  5.07 \pm 0.23 & 4.82 \pm 0.19      
\\
 b_{(2)}^{1,{\bf 8}}
&  & 0.66 \pm 1.49  &  0.66 \pm 1.18 &  & &   &
&    &   
\\
 b_{(3)}^{1,{\bf 8}}
&  & -0.25 \pm 0.95   & -1.14 \pm 0.82 &   &  &   &
&     &    
\\
 c_{(3)}^{1,{\bf 8}}
&  & 4.18 \pm 1.71  & 4.33 \pm 1.89 &  &   &  &
&    &    
\\
 \bar a_{(1)}^{1,{\bf 8}}
&  &  &   &   &  &  1.60 \pm 0.42  & 1.54 \pm 0.37     
& 1.57 \pm 0.44 &  1.49 \pm 0.33      
\\
 \bar b_{(2)}^{1,{\bf 8}}
&  &   &   &   &   &  &
&    &    
\\
 \bar b_{(3)}^{1,{\bf 8}}
&  &  &   &   &   &  &
&      &     
\\
 \bar c_{(3)}^{1,{\bf 8}}
&  &  &   &   &   &  &
&     &     
\\
 c_{(2)}^{1,{\bf 10}+{\bf \overline 10}}
&  &  &   &   &  &  &
& 0.37 \pm 0.52 &  0.71 \pm 0.40      
\\
 c_{(3)}^{1,{\bf 10}+{\bf \overline 10}}
&   &    &   &   &   &  &
&    &      
\\
 c_{(1)}^{1,{\bf 1}}
&  &  &   &   &   &  &
&   &     
\\
 c_{(2)}^{1,{\bf 27}}
&  &  &   &   &   &  &
&   &     
\\
\hline
\end{array}
\end{eqnarray*}
\end{table*}

\begin{table*}[htbp]
\caption{Fits to Experimental Data.}
\smallskip
\label{tab:fits2}
\begin{eqnarray*}
\begin{array}{|c|c|c|c|}
\hline
& {\rm Fit\ E} & {\rm Fit\ E^\prime} &  {\rm Fit\ F} 
\\
\hline
 p
& 2.72 & 2.71 & 2.70
\\
 n
& -1.91 & -1.92 & -1.93
\\
 \Lambda
& -0.61 & -0.59 & -0.59
\\
 \Lambda \Sigma^0
& -1.49 & -1.51 & -1.53
\\
 \Sigma^+
& 2.45 & 2.45 & 2.46
\\
 \Sigma^0
& 0.64 &  0.64 & 0.65
\\
 \Sigma^-
&  -1.16 & -1.16  &  -1.15 
\\
 \Xi^0
& -1.26  & -1.27 &  -1.27
\\
 \Xi^-
& -0.64 & -0.65  &  -0.65
\\
 \Delta^{++}
& 5.64 & 5.47 &  6.14
\\
 \Delta^{+}
& 2.67 & 2.58 & 2.79 
\\
 \Delta^{0}
& -0.30 & -0.30 &  -0.56
\\
 \Delta^{-}
& -3.28 & -3.19 &  -3.91
\\
 \Sigma^{*+}
& 2.97 & 2.95 &  3.49
\\
 \Sigma^{*0}
& 0.05 & 0.08 &  0.10
\\
 \Sigma^{*-}
& -2.86  & -2.80 & -3.28
\\
 \Xi^{*0}
& 0.41 & 0.46 &  0.77
\\
 \Xi^{*-}
& -2.45 & -2.41 &  -2.65
\\
 \Omega^-
& -2.03  & -2.02 & -2.02 
\\
 {1 \over \sqrt{2}} p\Delta^+
& 2.48  & 2.48 &  2.48
\\
 {1 \over \sqrt{2}} n\Delta^0
& 2.48 & 2.48 &  2.48
\\
 {1 \over \sqrt{2}} \Lambda \Sigma^{*0}
& 2.07  & 2.09 & 1.94 
\\
 {1 \over \sqrt{2}} \Sigma\Sigma^{*+}
& 2.11  & 2.09 &  2.28
\\
 {1 \over \sqrt{2}} \Sigma\Sigma^{*0}
& 0.96 & 0.95 &  1.42
\\
 {1 \over \sqrt{2}} \Sigma\Sigma^{*-}
& -0.19 & -0.19 &  0.56
\\
 {1 \over \sqrt{2}} \Xi\Xi^{*0}
& 2.07 & 2.09 &  2.30
\\
 {1 \over \sqrt{2}} \Xi\Xi^{*-}
& -0.19 & -0.19 &  0.56
\\
%
%%%
%
 \sigma^{\rm theory}
& 0.024 & 0.035 & 0.06 \\
 {\rm dof}
& 5 & 2 & 1 \\
 a_{(1)}^{1,{\bf 8}}
& 4.48 \pm 0.07  & 4.53 \pm 0.10 & 4.44 \pm 0.14 
\\
 b_{(2)}^{1,{\bf 8}}
& 0.78 \pm 0.13   &  0.73 \pm 0.18 &  0.65 \pm 0.33
\\
 b_{(3)}^{1,{\bf 8}}
& -0.09 \pm 0.14  & -0.16 \pm 0.19 &  0.21 \pm 0.32
\\
 c_{(3)}^{1,{\bf 8}}
& 3.96 \pm 0.38  & 3.88 \pm 0.46 &  6.38 \pm 2.18
\\
 \bar a_{(1)}^{1,{\bf 8}}
& 1.23 \pm 0.12  & 1.32 \pm 0.17 &  1.16 \pm 0.27
\\
 \bar b_{(2)}^{1,{\bf 8}}
&    &   &  0.13 \pm 0.55
\\
 \bar b_{(3)}^{1,{\bf 8}}
&     &   &  0.60 \pm 0.71
\\
 \bar c_{(3)}^{1,{\bf 8}}
&     &   &  -5.46 \pm 5.55
\\
 c_{(2)}^{1,{\bf 10}+{\bf \overline 10}}
& 0.34 \pm 0.06  & 0.34 \pm 0.08 &  0.34 \pm 0.14
\\
 c_{(3)}^{1,{\bf 10}+{\bf \overline 10}}
&     &  & -3.83 \pm 5.05 
\\
 c_{(1)}^{1,{\bf 1}}
& 0.03 \pm 0.03 & 0.05 \pm 0.04 &   0.07 \pm 0.07
\\
 c_{(2)}^{1,{\bf 27}}
& 0.10 \pm 0.10 & 0.02 \pm 0.16 &  - 0.06 \pm 0.25
\\
\hline
\end{array}
\end{eqnarray*}
\end{table*}

\end{document}